\documentclass[a4paper,11pt]{article}

\usepackage{siunitx}
\usepackage{jinstpub} 
\usepackage{mathtools}
\usepackage{setspace}
\usepackage{amsmath}

\author{The XENON collaboration: }
\author[a]{E.~Aprile,}
\author[b,c,1]{J.~Aalbers, \note{Now at: SLAC National Accelerator Laboratory, Menlo Park, CA 94025-7015, USA and Kavli Institute for Particle Astrophysics and Cosmology, Stanford University, Stanford, CA 94305-4085 USA}}
\author[d]{K.~Abe,}
\author[e]{F.~Agostini,}
\author[f]{S.~Ahmed Maouloud,}
\author[g]{L.~Althueser,}
\author[f]{B.~Andrieu,}
\author[h]{E.~Angelino,}
\author[b]{J.~R.~Angevaare,}
% ...
\affiliation[a]{Physics Department, Columbia University, New York, NY 10027, USA}
\affiliation[b]{Nikhef and the University of Amsterdam, Science Park, 1098XG Amsterdam, Netherlands}
\affiliation[c]{Oskar Klein Centre, Department of Physics, Stockholm University, AlbaNova, Stockholm SE-10691, Sweden}
\affiliation[d]{Kamioka Observatory, Institute for Cosmic Ray Research, and Kavli Institute for the Physics and Mathematics of the Universe (WPI), University of Tokyo, Higashi-Mozumi, Kamioka, Hida, Gifu 506-1205, Japan}
\affiliation[e]{Department of Physics and Astronomy, University of Bologna and INFN-Bologna, 40126 Bologna, Italy}
\affiliation[f]{LPNHE, Sorbonne Universit\'{e}, CNRS/IN2P3, 75005 Paris, France}
\affiliation[g]{Institut f\"ur Kernphysik, Westf\"alische Wilhelms-Universit\"at M\"unster, 48149 M\"unster, Germany}
\affiliation[h]{INAF-Astrophysical Observatory of Torino, Department of Physics, University  of  Torino and  INFN-Torino,  10125  Torino,  Italy}
% ...

\emailAdd{j.angevaare@nikhef.nl}
\emailAdd{alexey.elykov@kit.edu}
\emailAdd{darryl.masson@physik.uni-freiburg.de}
\emailAdd{stefano.mastroianni@na.infn.it}
\emailAdd{xenon@lngs.infn.it}
\usepackage{lineno}

\newcommand{\mv}{muon veto}
\newcommand{\nv}{neutron veto}
\newcommand{\db}{MongoDB database}

\newcommand{\order}[1]{$\mathcal{O}\left(#1\right)$}
\newcommand\iso[2]{\textsuperscript{#2}#1}
\newcommand{\reg}[1]{\textsuperscript{\small{\textregistered}} }

\DeclareSIUnit{\byte}{B}
\DeclareSIUnit{\lightspeed}{c}
\DeclareSIUnit{\year}{yr}
\DeclareSIUnit{\pe}{PE}
\DeclarePairedDelimiterXPP\BigOSI[2]%
  {\mathcal{O}}{(}{)}{}%
  {\SI{#1}{#2}}

\title{The Triggerless Data Acquisition System of the XENONnT Experiment}

\abstract{ 
The XENONnT detector uses the latest and largest liquid xenon-based time projection chamber (TPC) operated by the XENON Collaboration, aimed at detecting Weakly Interacting Massive Particles and conducting other rare event searches. 
The XENONnT data acquisition (DAQ) system constitutes an upgraded and expanded version of the XENON1T DAQ system. 
For its operation, it relies predominantly on commercially available hardware accompanied by open-source and custom-developed software. 
The three constituent subsystems of the XENONnT detector, the TPC (main detector), \mv, and the newly introduced \nv, are integrated into a single DAQ, and can be operated both independently and as a unified system.
In total, the DAQ digitizes the signals of 698 photomultiplier tubes (PMTs), of which 253 from the top PMT array of the TPC are digitized twice, at $\times10$ and $\times0.5$ gain.
The DAQ for the most part is a triggerless system, reading out and storing every signal that exceeds the digitization thresholds. 
Custom-developed software is used to process the acquired data, making it available within $\BigOSI{10}{\s}$ for live data quality monitoring and online analyses.
The entire system with all the three subsystems was successfully commissioned and has been operating continuously, comfortably withstanding readout rates that exceed \SI[per-mode=symbol]{\sim 500}{\mega\byte\per\second} during calibration.
Livetime during normal operation exceeds $99\%$ and is $\sim90\%$ during most high-rate calibrations.
The combined DAQ system has collected more than 2 PB of both calibration and science data during the commissioning of XENONnT and the first science run. 
\\
}


\begin{document}
\maketitle
\flushbottom

\section{Introduction}
A variety of experiments use time projection chambers (TPCs) filled with liquid noble elements (usually xenon or argon) in the search for Weakly Interacting Massive Particle (WIMP) dark matter and rare radioactive decays~\cite{1t_instrument,lz,pandax,nexo,darkside}.
While the details of each detector differ, common design features include arrays of photosensors at the ends of the drift region and accompanying readout systems.

Interactions in a dual-phase TPC are observed via two processes: scintillation and ionization.
When a particle interacts with either the electrons or nucleus of a target atom, prompt scintillation light and liberated electrons are produced, resulting in two signals.
Two arrays of photosensors, photomultiplier tubes (PMTs) in XENONnT, are located above and below the cylindrical drift region to capture these signals.
The detected scintillation light is referred to as the ``S1'' signal, while the electrons are drifted under an external electric field towards the liquid-gas interface.
When the electrons reach this interface, a much stronger electric field extracts them from the liquid and causes electroluminescence in the gas, producing additional proportional scintillation that is detected and referred to as the ``S2'' signal.
The time between the S1 and S2 signals, which is the drift time of electrons in the liquid phase, as well as the pattern of illumination on the top PMT array caused by the S2, are used to reconstruct the interaction vertex in the detector.
The S2 is typically much larger than the S1, and the relative sizes of these two signals are used to discriminate between electronic recoil (ER) and nuclear recoil (NR) interactions.

The XENON collaboration has operated a series of increasingly larger dual-phase xenon TPCs at the INFN Laboratori Nazionali del Gran Sasso (LNGS) in central Italy for nearly two decades, probing WIMP-nucleon cross-sections down to \SI{4.1e-47}{\square\cm} (for a \SI[per-mode=symbol]{30}{\giga\electronvolt\per\lightspeed\squared} WIMP)~\citep{1tonneyear}.
The latest is the TPC of the XENONnT detector, containing \SI{5.9}{\tonne} in its active target mass and is expected to be sensitive to spin-independent WIMP-nucleon cross-sections down to \SI{1.4e-48}{\square\cm} ({for a \SI[per-mode=symbol]{50}{\giga\electronvolt\per\lightspeed\squared}} WIMP)~\cite{nt_mc}.

\section{From XENON1T to XENONnT}
The upgrade from XENON1T to XENONnT saw the TPC increase in size from \SI{\sim 1}{\meter} in diameter and length to \SI{\sim 1.3}{\meter} and \SI{\sim 1.5}{\meter}, respectively, to accommodate a larger target mass.
This increase in TPC size was accompanied by a corresponding increase in the number of PMTs to 494, 253 in the top array in the gas phase and 241 in the bottom array in the liquid below the target.
This constitutes a two-fold increase from XENON1T where the TPC was instrumented with 248 PMTs.

As detectors continue to grow in size, the maximum drift time (the drift length) of electrons in the TPC grows in accordance.
A lower drift field of \SI[per-mode=symbol]{23}{\volt\per\cm} in the first science run (SR0) of XENONnT~\citep{low_er_nt} compared to \SIrange[range-phrase=--, per-mode=symbol]{81}{120}{\volt\per\cm} in XENON1T~\citep{1tonneyear} means a further increase in the drift length.
The need to store and read out one continuous drift length of data, which can exceed \SI{2}{\milli\second}, is alleviated by the firmware used by most of the readout hardware.
This digital pulse processing with dynamic acquisition windows (DPP-DAW) firmware was developed in collaboration with CAEN~\citep{caen_general} for XENON1T~\citep{1t_daq}, and an updated version was used for XENONnT.
It affords many useful techniques such as baseline suppression or zero length encoding (ZLE), dynamically-sized acquisition windows that automatically extend as long as the input is above the digitization threshold, and the independent and continuous readout of each channel.
The digitization thresholds are set relative to dynamically-calculated baselines.
However, increased drift time leads to an increased temporal width of S2 signals as the freed electrons diffuse over a larger amount of time.
This increased temporal width directly increases data rates as signals remain above threshold for longer.
Thus, new challenges arise as drift times increase and S2s become longer.

A new active \nv{} sub-detector was built to suppress the NR background from radiogenic neutrons generated through spontaneous fission and alpha-neutron reactions, as these mimic WIMP-induced signatures.
It is made of an octagonal structure (\SI{3}{\meter}-high and \SI{4}{\meter}-wide) placed inside the water tank around the cryostat that houses the TPC and is optically separated from the existing muon veto~\citep{nt_mc}.
To improve the neutron detection efficiency, the water will be loaded with gadolinium (Gd).
A total of 120 Hamamatsu 8'' high quantum efficiency PMTs with low-radioactivity windows are placed along the lateral walls.
Neutrons that leave the TPC volume are moderated by the water around the cryostat before being captured Gd or H.  
A gamma-ray cascade with total energy of about 8 MeV is generated for capture of Gd and a single 2.2 MeV gamma is emitted in the case of capture on H.
The gammas in water, mainly through Compton scattering, are converted into electrons and ultimately into Cherenkov photons.
Monte Carlo studies indicate that the \nv{} is expected to reduce the total NR background by a factor of six~\cite{nt_mc}. 
Dedicated hardware was implemented to manage the \nv{} data readout.
The water Cherenkov muon veto surrounding the cryostat, instrumented with 84 PMTs, is otherwise largely unchanged from XENON1T~\cite{1t_daq, mv_paper}.

\subsection{General DAQ upgrades}\label{sec:general}

\begin{figure}[ht]
\includegraphics[width=\textwidth]{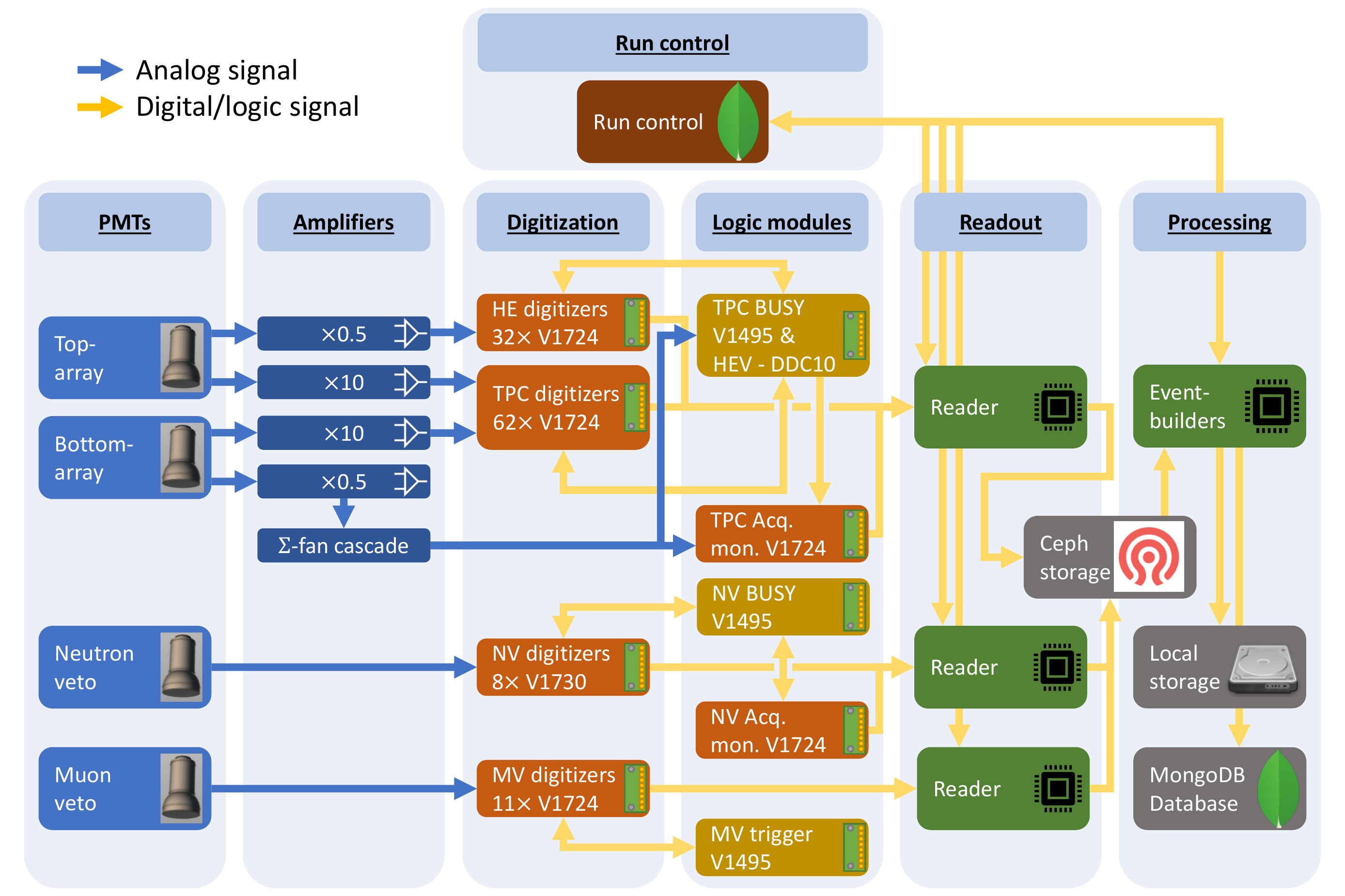}
\caption{
    The XENONnT DAQ layout at several stages.
    The TPC PMT signals are amplified before being digitized by CAEN V1724 modules. 
    The \nv{} (NV) and \mv{} (MV) PMT signals are digitized by the associated V1730S and V1724 digitizers.
    The  top array is digitized twice at $\times$10 and $\times$0.5 gain, the signals from the latter constitute the high energy system. 
    The $\times$0.5 gain signals of the bottom array are fed into the sum-signal fan cascade ($\Sigma$-fan cascade).
    The busy (V1495) and acquisition monitor (V1724) modules handle the busy logic and monitor the system performance.
    The reader servers read out the digitizers and write the data to a common (Ceph) storage disk. 
    The eventbuilder servers do the processing of the reader data which get written to local storage where it can be distributed to other storage sites. 
    A portion of the processed data is also written to the MongoDB database where it can be accessed for online data monitoring.
    The control of the DAQ is achieved through the website that communicates commands (run control) to the readers/eventbuilders.}
\label{fig:overview}
\end{figure}

The XENONnT data acquisition (DAQ) system is an evolution of that which was successfully used for XENON1T~\cite{1t_daq}.
Many aspects of the system have received modifications and improvements based on the XENON1T system and the experience of operating it.
An overview of the new system design is shown in~\autoref{fig:overview}.

One challenge of the general DAQ design is the range of sizes and shapes of signals the system must handle.
The \mv{} and \nv{} are Cherenkov detectors, registering photon signals over an interval of at most $\BigOSI{1}{\micro\second}$.
The TPC, in contrast, must record both S1 and S2 signals.
S1s can be very small, down to a single photon, and are very fast, lasting up to $\BigOSI{100}{\ns}$.
S2s are much larger, potentially millions of photons, and can have temporal widths exceeding \SI{100}{\us}.
Representative signals in all three subsystems are shown in~\autoref{fig:tpc_signal}.

While XENON1T used the Phillips Scientific 776 amplifiers with dual $\times$10-gain outputs, XENONnT uses custom dual-gain ($\times$10 and $\times$0.5) amplifiers developed at the University of Zurich~\cite{julien_amp_thesis}.
The low-gain signals from the top PMT array are digitized separately from the high-gain signals to try to improve energy and position reconstruction involving large signals that otherwise saturate the input stage of the digitizers.
The low-gain signals from the bottom array are summed together and used by the high-energy veto, discussed in~\autoref{sec:busyhev}.
This results in the XENONnT TPC effectively having three times the number of PMT readout channels of the XENON1T TPC (747 compared to 248), as the number of PMTs is doubled and half are read out twice.

\subsection{Triggerless data streams}

The XENON1T TPC utilized a triggerless readout and a central buffer built from a MongoDB database~\cite{mongo}.
Triggering software ran live over this database and determined ``events'' which were written to disk for later analysis, the remainder of the data was deleted.
While this paradigm was successful at realizing a very low effective trigger threshold, the estimation of some backgrounds was more difficult due to the forced truncation of events after a certain maximum duration.
Additionally, this database would not scale effectively to match the increased load foreseen by the demands of a larger system.
The solution is to forego the software trigger and save all the data, leaving the determination of events to much later in the data processing pipeline.


The removal of all triggers except the per-channel digitization threshold does not lead to significantly increased storage requirements.
The primary driver of data rate is not PMT dark counts or other small signals seen by only a few channels, rather it is large S2 signals that can remain above threshold for
\order{10 - \SI{100}{\micro\s}}, and are seen by a large number of channels.
For instance, in typical conditions during SR0, S2s from single electrons account for $30\%$ of all reconstructed signals but only $2\%$ of the data volume.
In contrast, for very large S2s these values are reversed, accounting for $2\%$ of the reconstructed signals but $30\%$ of the data volume.
Any trigger would be configured to save the large S2s, so additionally saving everything else (mostly S2s) does not represent a significant increase on the requirements of long-term storage.    
To support this, work was done studying data formatting and data compression, and a storage format was chosen that compresses more efficiently than the XENON1T storage format.
Further, the removal of the software trigger eliminates the requirement for a database that can act as a base for fast triggering software, so the readout processes write data directly to high-speed disks in a continuous stream.

\subsection{Fast data processing \& immediate data availability}
In addition to the hardware upgrades from XENON1T to XENONnT, the readout and processing software was also upgraded. 
To handle the continuous data stream of roughly three times the number of channels, the processing framework \textit{PAX} \citep{pax} was replaced by the generic framework \textit{strax}~\citep{strax} and implemented for XENONnT in \textit{straxen}~\citep{straxen}.
Strax and straxen are referred to as strax for simplicity. 
Strax is written in python and was initially based on a re-write of PAX with a different memory model.
It uses packages from the scipy stack~\citep{numpy, scipy}, just-in-time compilation (numba)~\citep{numba}, and a tabular data format to allow for fast processing by exploiting autovectorization.
While PAX achieved processing speeds of \order{\SI{100}{\kilo\byte\per\second\per\text{core}}}, strax can process data at rates of \order{10-\SI{100}{\mega\byte\per\second\per\text{core}}}.
Strax achieves its highest per core processing speeds when running on only a few cores but also allows parallelization to tens of cores, albeit at lower per core performance.
For the data rates observed during SR0, including the associated calibration periods where a higher rate was expected, the processing time was much lower than the data collection time. 
The strax framework will be further elaborated on in \autoref{sec:live_processing}. 

Strax does complete online reconstruction of all the data within $\BigOSI{10}{\second}$ after the PMTs detect light.
This allows the detector performance and stability to be monitored with high-level data without the need for selections or triggers. 
To enable remote, online access to the data while it is being processed at the DAQ, several types of data are uploaded to the \db~in a dedicated collection. 
This database is accessible from outside LNGS, such that it can be retrieved from anywhere.
During normal operation, these data are available online within $\BigOSI{30}{\second}$. 
With online data access the performance can be monitored using fully reconstructed data.
This is especially useful for stability checks, as well as detailed feedback on operations with rapidly changing conditions such as calibrations or changing field configurations.
These online data are blinded, and only when purposely unblinded and reprocessed the science results~\citep{low_er_nt} are obtained.

\subsection{Neutron veto DAQ}
The goal of the neutron veto in XENONnT is to detect the capture process of those neutrons responsible for NR background events, which can mimic the interaction of a WIMP. 
A neutron detection tagging efficiency greater than 85\% is desired~\citep{nt_mc}. 
Since the expected Cherenkov signal in the case of neutron capture by H is of about \SI{20}{PE} in total, it is important to have a high detection efficiency for each photon.
To achieve such a high efficiency in a trigger-based DAQ architecture it would have been necessary to reduce the number of coincident PMTs that form the trigger.
This, in turn, would have lead to an increased number of triggers and acquired data, making it challenging for the DAQ readout.

Therefore, the \nv{} DAQ is designed around a triggerless data collection scheme like the TPC. 
Its ability to provide both the pulse shape and the timestamp of each PMT signal supports data collection with fully independent channels without the use of a global trigger, typically based on channel multiplicity.
As will be described in~\autoref{sec:live_processing}, the event building is done in software after data acquisition, where timestamps and coincidences between PMT signals are used to define events. 
This architecture based on a readout system of independent channels allows the acquisition of all the PMT signals above the digitization threshold and the lowering of the energy threshold.

PMT characteristics such as dark rate, afterpulsing and timing resolution are essential for the choice of front-end electronics. 
In particular, the dark rate puts a limit on the detection of very small signals, and can be used to estimate the accidental coincidence rate with a defined number of PMTs within a specific time window.
Operating with a threshold of 0.5 photoelectrons (PE), the measured PMT dark rate during detector commissioning was about \SI{0.96}{\kHz}, generating an accidental coincidence rate that exceeded \SI{4}{\kilo\hertz} for a 2-fold coincidence between two random \nv{} PMTs. 
In addition, the materials in the sub-detector itself (PMTs and stainless steel structure) induce events that mimic NR signals in the \nv{} of $\BigOSI{100}{\hertz}$.

To efficiently tag neutron events, the electronics must be able to acquire signals ranging between 0.5~PE and $\BigOSI{100}{\pe}$, requiring a wide dynamic range. 
The small signals last about \SI{100}-\SI{200}{\ns}, resulting mainly from the dark rate, and define the lower limit for the \nv{} data throughput. 
In contrast, gamma and beta particles from materials radioactive decays with a typical rate of $\BigOSI{100}{\hertz}$ exhibit waveforms that last up to \SI{10}{\us} (considering signals and associated afterpulses) in many channels, requiring a much higher data collection rate. 
Therefore, the readout electronics must be able to handle extremely different time acquisition windows with the presence of sharp peaks of data rate.

In addition, the fast response (few ns) of PMTs used by the \nv{} to acquire single photoelectrons requires a fast waveform digitizer for signal sampling. 
This high time resolution is necessary to efficiently separate neutrons produced close to and far from the TPC cryostat, where the former are the primary target.

\subsection{Three integrated DAQ subsystems}

One requirement for the XENONnT DAQ system was that the three DAQ subsystems (TPC, \mv, and \nv) should be able to operate both independently and as one combined system.
While both the TPC and \mv{} in XENON1T used the same \SI{50}{\mega\hertz} clock signal, there was no synchronization of the start signals issued to the two readout systems, so there was some variation in reconstructed timestamps between the two detectors, and data from the two subsystems were analyzed separately.
The trigger signal of the muon veto was recorded in one of the TPC's digitizers, but this did not provide the equivalent timestamp in the muon veto's data stream, thus viewing the corresponding event as observed in the muon veto required analysts to perform additional steps.

To ameliorate this, the XENONnT DAQ was designed to allow for the start signal from one subsystem to be issued directly to one or both of the others, essentially combining them into one and ensuring that timestamps recorded in one can be directly compared to those from another.
Subsystems can be combined or ``linked'' together as determined by the requirements of the data being taken, or can operate independently.
Subsystems operating in linked modes are controlled as a single operational unit, and the data they record are combined at the readout level and processed together to facilitate handling and analysis.

\begin{figure}[t!]
\includegraphics[width=1\textwidth]{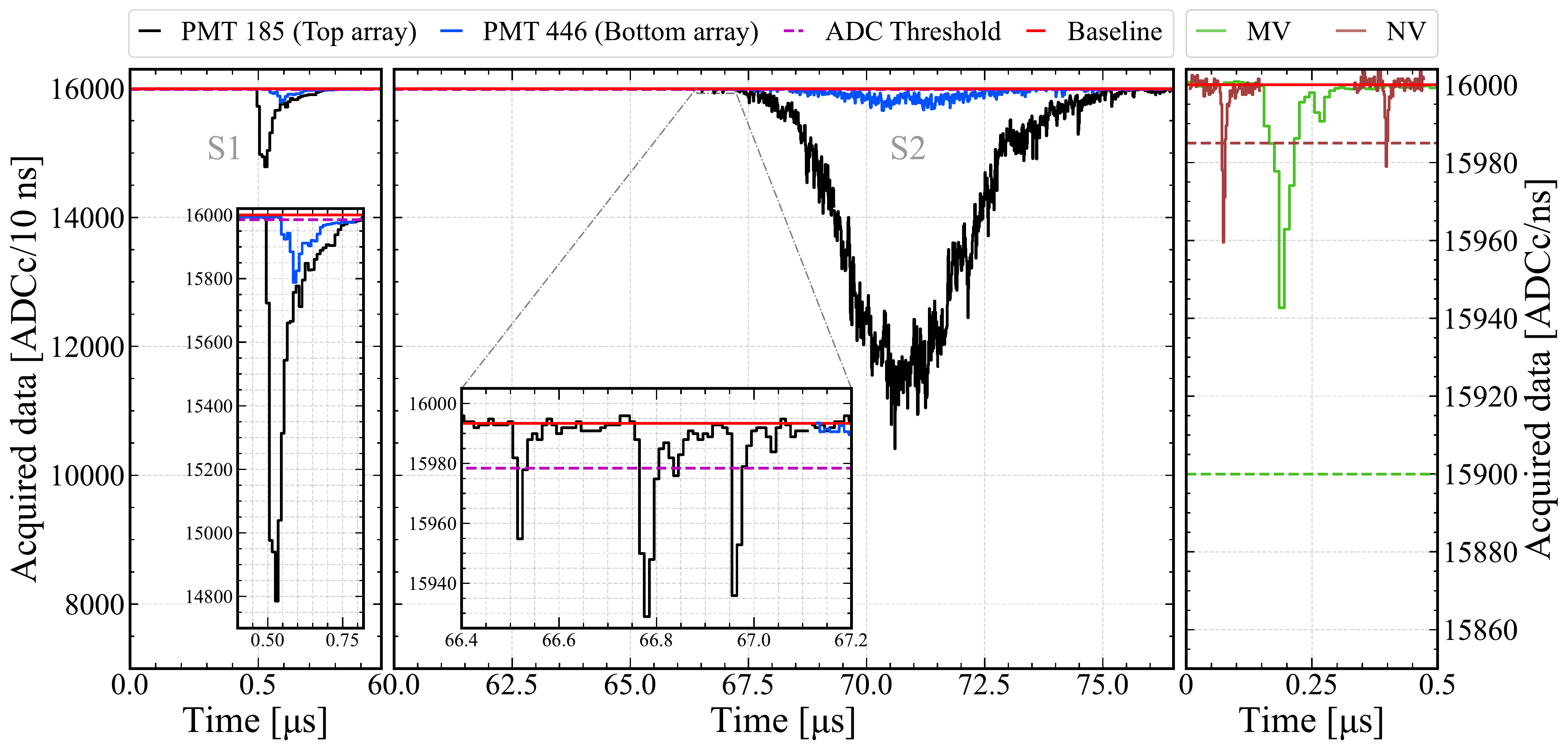}
\caption{
    An illustration of the variety of signals read out by the XENONnT DAQ in ADC counts (ADCc). 
    The left and middle panels showcase raw signals from a selection of two TPC PMTs in a single event. 
    The inset in the left panel zooms in on the S1 to emphasize its narrow width and short risetime, contrasting with the wide S2 in the middle panel. 
    An inset in the middle panel zooms in on minute signals in the leading edge of the S2 waveform. 
    In both panels, the black lines correspond to signals from a PMT in the top array, and the blue from the bottom array. 
    The red and purple dashed lines represent the baseline and digitizer threshold, respectively. 
    The rightmost panel shows signals that were recorded by individual channels in the \mv~(green) and \nv~(dark red) DAQ subsystems.
    These signals are not correlated with the ones showcased for the TPC channels.
    Typical \mv~ and \nv~ thresholds are depicted with dashed lines of matching color.
    The higher sample rate of the digitizers employed by the triggerless \nv~subsystem is clearly visible in comparison to the triggered \mv.
    In this case, a relative baseline is shown in red for illustration purposes only.
    As explained in \autoref{sec:mv_hardware}, this \mv{}~signal is read out, despite not being above threshold, because of the hardware coincidence trigger during this time interval.
    } 
\label{fig:tpc_signal}
\end{figure}

\section{Data Acquisition \& Readout}

The data acquisition is organized in two broad schemes following the system used in XENON1T.
At the greatest scope are so-called ``science runs'' which represent a period of months or even years with a targeted science objective where the detector conditions are held constant.
For the daily operation of the experiment, the organizational unit called ``runs'' is used, where each run represents a continuous period of a few minutes up to a few hours using a set of configuration options that remain constant for the duration of the run.

The three DAQ subsystems rely predominantly on commercially-available analog front-end electronic modules supported by custom hardware.
The firmware and software include both custom components and some provided by CAEN.
All commercially available CAEN products are marked with their model number throughout this work, the reader is referred to the company's website for more details and manuals~\citep{caen_general}.
All DAQ hardware is installed within eight racks located in the DAQ room on the first floor of the XENON service building in Hall B of LNGS.
An air conditioning system provides cooling for the electronics, maintains a constant air temperature in the room, and reduces the collection of dust.

\subsection{Analog electronics}

The PMTs of the TPC are powered by an array of multi-channel CAEN A7030LN, A1536LN and A1535LN high-voltage supplies. 
Five CAEN A7435SP high-voltage (HV) boards supply power to the \nv{} PMTs, and four CAEN A1535SP boards power the PMTs of the \mv{}.
The above HV boards are housed in separate CAEN SY4527 Universal Multichannel Power Supply System crates.
HV boards are known to produce high-frequency switching noise.
Hence, before being supplied to TPC PMTs the output from each HV board is passed through a custom filter box.
Within each filter box, every HV channel line goes through a low-pass filter, removing electronic noise with frequencies greater than \SI{\sim 250}{\kilo\hertz}.

Before installation, all TPC signal cables were grouped based on the location of their corresponding PMTs in the arrays and assigned to specific hardware modules.
The resulting cable map ensures an equal distribution of the data load on readout electronics and was used as a guide throughout the hardware installation process.
Signals from both PMT arrays are passed through custom amplifiers as mentioned in~\autoref{sec:general}.
As illustrated in \autoref{fig:overview}, both the high- and low-gain signals from the top PMT array are propagated to dedicated groups of digitizers.
However, only high-gain signals from the bottom PMT array are passed to digitizers.
The low-gain signals from the bottom array are summed up using a cascade of linear fan-in/fan-out modules and are used by the high energy veto (HEV).
Lastly, each DAQ subsystem hosts a range of logic modules that are used for distributing initialization and trigger signals.

\subsection{Digital electronics}
\label{sec:digi_ele}

For time synchronization across its subsystems the XENONnT DAQ relies on a CAEN DT4700 clock generator module.
Its \SI{50}{\mega\hertz} low-voltage differential signaling (LVDS) outputs are propagated via seven shielded custom-manufactured cables to the first digitizer in each VME crate. 
Shielding the clock-carrying cables reduces the amount of external noise that can be injected into the cables, improving the stability of the clock signal.
The propagated signals are then distributed within each VME crate by shorter clock cables from digitizer to digitizer, ensuring the temporal synchronization of the entire DAQ system.
Time offsets in these clock chains were manually calibrated out, securing synchronization well below the digitizer temporal resolution.

Additionally, a GPS timing module~\cite{gps_slave_master} is used to distribute a \SI{0.1}{\hertz} trigger signal to dedicated digitizers in each DAQ subsystem.
Each trigger is associated with a GPS timestamp (accurate to \SI{\sim 10}{\ns}), providing another layer of time synchronization within the DAQ. 
The same signal can also be used for absolute time synchronization with other experiments.

\subsubsection{TPC}

The core of the TPC readout is formed by 95 CAEN V1724 digitizers running the DPP-DAW firmware, an updated version of what was used in XENON1T~\cite{1t_daq}.
The V1724 is an 8-channel board featuring a sample rate of \SI{100}{\mega\hertz}, a dynamic input range of \SI{2250}{\milli\volt} (input impedance \SI{50}{\ohm}) with 14-bits of resolution, and an input bandwidth of \SI{40}{\mega\hertz}.
Of these digitizers, 62 read the 494 high-gain signals from the top and bottom arrays, 32 read the 253 low-gain signals from the top PMT array, and 1 acts as the TPC's Acquisition Monitor detailed in~\autoref{sec:acqmon}.
The boards are distributed across five VME crates and connected to readout servers via daisy-chained optical links.
Most optical links contain the maximum of 8 digitizers, while the acquisition monitor is read out via its own dedicated optical link to ensure it never goes busy.

One CAEN V2718 crate control module is used as a synchronizing module to produce the sync/start/stop digital input (S-IN) signal that begins and ends the acquisition.
This signal is distributed to all digitizers via logic fans, with the signals all reaching their respective digitizers with a spread of $<$\SI{4}{\nano\second}.
Additionally, this module provides gate logic signals that control the propagation of the S-IN signal to the muon and neutron veto digitizers that are activated during linked-mode operation.
A periodic external trigger signal can be generated by this module, which is distributed both to all digitizers and an external LED pulser used to calibrate the response of the PMTs.
The third type of module is a general-purpose CAEN V1495 board running custom firmware, which manages the busy subsystem detailed in~\autoref{sec:busyhev}.

Finally, two NIM crates hold the logic fan modules used to distribute the S-IN and trigger/veto signals to all TPC digitizers, as well as a gate module, a NIM-TTL level converter, and a delay generator.
These latter two are used to connect and synchronize the TPC DAQ with the LED calibration system.

\subsubsection{Muon veto\label{sec:mv_hardware}}

The muon veto readout is unchanged from XENON1T as described in~\cite{1t_daq}, though some additional connections were made between this subsystem and those of the TPC and \nv.
Eleven V1724 digitizers with the default ZLE firmware form the readout system, although the zero length encoding features are not used.
Three optical fibers are used to read out these digitizers.
A V2718 module provides the S-IN signal for these digitizers during unlinked operation.
A CAEN V976 unit serves as a logic fan to distribute both this S-IN signal and that of the TPC during linked operation to all muon veto digitizers.
A V1495 board acts as a programmable trigger unit, allowing the user to specify both the number of participating channels and the coincidence window necessary to generate a hardware trigger.

\subsubsection{Neutron veto}
The 120 PMTs of the \nv{} are connected to the readout electronics and HV system located in the DAQ room by means of \SI{30}{\m} coaxial cables with separate grounding for signal and high voltage cable lines. 
A custom-made patch panel mounted on the back side of the \nv{} rack gathers HV lines in one section and signal lines in another.
Signal lines are directly connected to the front-end electronics via a panel feedthrough.
HV lines are low-pass filtered to reduce high frequency noise ($\gtrsim$ MHz) and connected to the CAEN A7435SP HV boards.

To take advantage of the fast response of the PMTs and to efficiently reconstruct the fast component of Cherenkov photons in the \nv{} sub-detector, eight CAEN V1730S new generation digitizers are used to acquire PMT signals.
Each V1730S board is a VME 6U module housing a 16-channel 14-bit \SI{500}{\mega\hertz} flash ADC.
The input dynamic range can be set to either \SI{2}{\V} or \SI{0.5}{\V} on single ended MCX coaxial connectors.
During commissioning and SR0, the \SI{2}{\V} dynamic range was used.
The input section is \SI{50}{\ohm}-coupled and feeds a programmable gain amplifier to select the suitable analog range.
In case all the buffer memory is filled, a busy condition occurs and a logic module inhibits the data acquisition for all the boards (as described in~\autoref{sec:busyhev}).
The V1730S digitizers are operated with the DPP-DAW firmware like the TPC digitizers.
An exemplary \nv{} waveform is shown in~\autoref{fig:tpc_signal}.

The V1730S module is also able to work with a common global trigger, either coming from the digital input external trigger (TRG-IN) input or a coincidence trigger. 
In particular, the external trigger mode is used by the \nv{} system during calibration.
A V2718 board hosted in the VME crate generates several control signals (mainly the start-of-acquisition and calibration signals) that are subsequently distributed to the digitizers via logic fan-in/fan-out modules.
Two additional boards are hosted in the \nv{} crate: a V1495 to manage the V1730S busy signals and provide the veto signal, and a V1724 digitizer that serves as an acquisition monitor.

The \nv{} digitizers are connected to a readout server via two optical links; one daisy-chains the V1730S digitizers while the second is for the V1724 acquisition monitor.
In order to synchronize all the digitizers in the \nv{} DAQ and limit the clock uncertainties to below \SI{\sim 1}{\ns}, an external common clock reference feeds all the modules. 
The V1724 digitizer receives the common \SI{50}{\MHz} clock signal (see~\autoref{sec:digi_ele}), which is then upconverted to \SI{62.5}{\MHz} via a phase locked loop device and propagated through the V1730S boards.
Lastly, several auxiliary electronic modules are hosted in a NIM crate, managing the distribution of calibration triggers and run start signals. 

\subsection{Busy \& high-energy veto}
\label{sec:busyhev}

The V1724 and V1730S digitizers have a limited on-board memory buffer for storing data between digitizing and readout, amounting to \SI{1}{\mega\byte}/channel and \SI{10.24}{\mega\byte}/channel, respectively.
If incoming data accumulates in the digitizer’s memory buffer faster than it is read out, the buffer will become full and the digitizer will no longer be able to acquire new signals, rendering it \textit{busy}.
To ensure the integrity of individual events the triggerless TPC and \nv{} DAQ subsystems employ a hardware-based \textit{busy veto}.
Each subsystem hosts a general-purpose V1495 board, equipped with field-programmable gate array (FPGA) firmware developed in-house.
When a digitizer enters the busy state it emits an LVDS signal via a pair of connectors on its front panel.
These signals are propagated via ribbon cables from each digitizer in each subsystem to its respective V1495 module. 

Whenever the V1495 recognizes that any of the digitizers emits an LVDS busy signal, it outputs a veto NIM signal for a fixed duration. 
This veto signal is distributed to all the digitizers within the subsystem, inhibiting data acquisition for \SI{1}{\ms} or until none of the digitizers are busy.
Within the FPGA firmware, busy intervals are assigned with start and stop NIM signals. 
These are also output from the V1495 board and propagated to the relevant acquisition monitor digitizer, as explained in \autoref{sec:acqmon}.
The TPC V1495 board has a more advanced version of this firmware. 
Besides being responsible for the busy veto, it is also capable of generating an artificial periodic veto, with user controlled duration and frequency.
During detector commissioning, the water tank was empty and the detector was not shielded from radiation in the experiment hall. This periodic hardware-induced veto allowed the TPC DAQ to handle high background rates in addition to taking \iso{Kr}{83m} calibration data.
Additionally, the V1495 module performs several other important functions. 
In the TPC DAQ it collects and propagates the HEV signal, and assigns it with start and stop NIM signals that are read by the TPC acquisition monitor. 
In both the \nv{} and TPC subsystems the V1495 board is also responsible for the propagation of the LED trigger to the digitizers during LED calibration. 

The TPC DAQ also employs a hardware veto to reduce the load on the system during acquisition of high-rate calibration data.
This HEV was developed based on a commercially-available multipurpose digital pulse processor DDC-10 from SkuTek~\cite{ddc10}.
It hosts a variety of chips and daughter cards on a BlackVME S6 motherboard, including a Spartan-6 FPGA and a \SI{100}{\MHz}, 10-channel, 14-bit ADC.
Mounted on the FPGA is custom-developed firmware whose main goal is to identify and veto high-energy S2 signals.
The HEV digitizes the analog sum signal from the low-gain TPC channels of the bottom array, determining the risetime, width and integral of acquired signals.
If any identified S2 exceeds predefined threshold parameters, the HEV issues a 3 ms veto NIM signal.
To provide the HEV module with enough time to make the veto decision, data readout is delayed within the TPC digitizers by \SI{10}{\us}.
The veto signal generated by the HEV is propagated to the TPC V1495 module, from where it is distributed to each TPC digitizer.

Drift field conditions in the TPC during SR0 produced broad S2 signals with widths greater than $\BigOSI{10}{\us}$, which were found to have the largest contribution to the DAQ rate.
It is difficult to identify and characterize the shape parameters of such signals within the \SI{10}{\micro\second} time limit.
Hence, the HEV firmware has an additional operation mode, whose purpose is to veto low-amplitude high-width S2 signals that might last for $\BigOSI{10}{\micro\s}$.
In this mode, if the HEV is not able to determine the width and the risetime of the signal within $\BigOSI{5}{\us}$
and the signal's amplitude is still above the HEV threshold it will consider the signal to be a high-width S2, and will issue a veto.
The aforementioned HEV operation modes can be utilized separately or run in parallel.
A schematic view of the hardware-based veto systems described above is shown in \autoref{fig:overview}.
The operation of the HEV results in raw data reduction at the readout stage of up to 40\%, depending on the utilized HEV settings.
Throughout SR0 the HEV was utilized during AmBe and \iso{Rn}{220} calibration data taking.

\subsection{Acquisition monitors}
\label{sec:acqmon}
The TPC and the \nv{} DAQ subsystems each host a dedicated V1724 digitizer, whose aim is to collect information about the status of the DAQ itself and the operation of its hardware veto modules.
Both the TPC and \nv{} acquisition monitors receive the start and stop NIM signals from their respective V1495 veto modules, which indicate the boundaries of busy veto intervals. 
Additionally, these digitizers also acquire the \SI{0.1}{\hertz} NIM synchronization signal from the GPS module~\cite{gps_slave_master}. 
Uniquely for the TPC, its acquisition monitor also digitizes the same analog sum waveform signal that is seen by the HEV module.
To prevent the TPC acquisition monitor from ever going busy a relatively high threshold of 100\,ADCc (ADC counts) or \SI{14}{\milli\volt} is set on this channel. 
Furthermore, acquisition monitors are also excluded from the busy veto distribution scheme.

Acquisition monitor data are read out identically to the rest of the digitizers and incorporated into the overall data processing chain.
These data are then used to diagnose the performance of the busy and HEV systems, and to determine the deadtime they induce.
The measured deadtime under several operational modes is discussed in \autoref{sec:deadtime}.
Moreover, the same data are used as a basis for a data quality cut. 
The cut removes any events that could be misreconstructed due to missing information as a result of their proximity to a busy or a HEV veto interval.
The cut decreases the livetime and is accounted for in the exposure rather than the cut acceptance~\citep{low_er_nt}.

\subsection{Servers \& software}
Five server computers are responsible for the readout of all the digitizers, three for the TPC and one each for the \mv~and \nv.
Two additional servers provide backup capacity.
The TPC readout servers each have four \SI{960}{\giga\byte} write-intensive solid state drives which are configured together as a Ceph cluster~\cite{ceph} to form a single high-speed buffer disk with approximately \SI{10}{\tera\byte} of capacity that is accessible from all servers within the DAQ network.
While replication is possible using Ceph, it is not necessary for a short-term buffer disk, so the configuration is equivalent to RAID0 (data striping) to provide the highest access speeds.
This buffer disk can sustain simultaneous read and write operations from multiple sources at rates exceeding \SI[per-mode=symbol]{1}{\giga\byte\per\second}.
Data are stored on the Ceph buffer from the start of acquisition until the live processing for that run has successfully concluded, which is typically one or two hours, so very little data are lost in the event of disk failure.
The combined data rate from the three subsystems during science data taking is approximately \SI[per-mode=symbol]{40}{\mega\byte\per\second}, so the disk can potentially buffer data for a considerable amount of time in case of issues in the live processing.

Each readout server is equipped with at least one CAEN A3818 PCIe interface card.
Each A3818 supports up to 4 optical fibers, with each fiber capable of daisy-chaining up to 8 digitizers and supporting a maximum data throughput of \SIrange[range-phrase= --,per-mode=symbol]{80}{90}{\mega\byte\per\second}.
Digitizers and optical links are distributed to approximately balance the load on each of the readout servers.

The readout servers all run the \textit{redax} software package~\cite{redax}, which copies data from the digitizers and transforms it from the digitizer-native format into one compatible with the strax data processing package~\cite{strax,straxen}.
Data are read from digitizers in block transfers via the CAENVMElib C++ library, where each optical fiber is read out exclusively by a dedicated readout thread.
A round-robin technique is used, where each board on an optical fiber is successively polled.
When a digitizer has data available for readout, block transfers are performed until all data stored on that digitizer have been read into the server's memory.
The readout threads then transfer data asynchronously to processing threads, where the binary format transformation is performed.
Each processing thread periodically compresses its buffered output data and writes it to the Ceph buffer in fixed-time intervals called chunks following the chunking paradigm in strax as described in~\autoref{sec:live_processing}.
Chunks are labeled with the name of the readout process, the chunk number, and also the ID of the thread that wrote that chunk, which acts as a unique identifier.
Additionally, redax is responsible for programming the digitizers in preparation for each run via configurations it obtains from a central database.
Redax also writes status snapshots to this database once per second, including quantities such as the current state of that instance of redax, the amount of data currently buffered in memory, and the data rate for each channel of each digitizer being read out.

Six additional servers, called the \textit{eventbuilders}, are responsible for the live processing.
In case of high data rates or unlinked operation, when the three DAQ subsystems run independently, multiple hosts can process data simultaneously.
For low data rates only a single host is required for the processing.
Three eventbuilders are PRIMERGY RX2540 M4 Fujitsu servers with two Intel\reg{}Xeon\reg{}Gold 6128 CPUs at \SI{3.40}{\GHz} and \SI{202}{\giga\byte} of RAM each.
Additionally, there are three backup PRIMERGY RX2540 M1 Fujitsu servers with two Intel\reg{}Xeon\reg{}E5-2660 v3 CPUs at \SI{2.60}{\GHz} and \SI{135}{\giga\byte} of RAM each. 
These backup servers were also used in XENON1T. 
Two of these three mainly serve as extra redundancy, while the third acts as a general purpose machine with access to the latest data.
This machine, for example, automatically produces online monitor plots and handles requests for retrieving the same (as explained in \autoref{sec:olmo}).

\section{Live Processing}

The data stream of raw data from the digitizers is fully processed onsite at LNGS.
The triggerless data stream is handled by the data stream processor, strax~\citep{strax,straxen}. 
Using live processing and online data storage, data can be accessed while their collection is still ongoing.

\label{sec:live_processing}
\subsection{Data stream versus discrete events}

The triggerless design of the XENONnT DAQ results in a continuous data stream. 
For processing as well as storage purposes, handling discrete time intervals of data is advantageous as it allows for parallelization.
To this end, the digitizer data which are read out by redax~\citep{redax} are partitioned in \SIrange[range-phrase= --]{5}{20}{\second} time intervals called chunks.
Each chunk is accompanied with an overlap region of \SI{\sim 0.5}{\s} to the previous and following chunk. 
As such, the overlap region is being saved twice, once with the previous chunk and once with the following. 
These overlap regions, called the pre- and post-chunk, are processed together with a chunk to ensure that each process has access to sufficient data to do the reconstruction.
This is important for reconstructing S1/S2 signals (peaks) whose data might otherwise be split into consecutive chunks.
Strax searches for time regions where there are no data for \SI{1}{\us} within the pre- and post-chunk and discards the data before (pre-chunk) or after (post-chunk) this time region.
This discarded time region will instead be processed together with the previous or next chunk. 
If no time interval of \SI{1}{\us} is found within the overlap region, artificial deadtime would be inserted, which was never required for the entire SR0 dataset including calibration data.
Due to this temporal separation between chunks, single chunks of low-level data are handled independently, allowing for processing in parallel.
For high level data, such as events, the processing is based on stateful algorithms which are therefore single threaded and may rearrange chunk boundaries.

\subsection{Strax(en) data format}

The processing at LNGS is handled by the eventbuilders and uses the publicly-available strax framework~\citep{strax,straxen}. 
Strax is a purely python-based streaming processor.
Autovectorization, just-in-time compilation, and a tabular data format make the processing fast. 
The tabular data format is achieved by fixing the shape of the data fields in software. 
At the level of PMT traces~(\autoref{fig:tpc_signal}), this is achieved by splitting one variable-length PMT trace into a sufficient number of fixed-length intervals.
The data are organized in a hierarchical structure of ``datatypes''.
At higher level datatypes, like S1/S2-peaks, the summed waveform of all PMTs is down-sampled to a fixed number of samples.

There are several steps in the processing, from PMT-traces as in \autoref{fig:tpc_signal} at the lowest level, to a fully reconstructed S1 and S2 pair originating from one physical interaction within the TPC at high level. 
The PMT-traces are stored as \textit{raw-records} as they are the lowest level (raw) datatype that is stored long term.
The S1s and S2s are saved as \textit{peaks} level data which can be grouped in time to form \textit{events}. 
The time scales of the typical objects in raw-records, peaks and events differ by orders of magnitude. 
For example, \autoref{fig:tpc_signal} shows that raw-records can be of $\BigOSI{1}{\us}$ and an S2 peak of $\BigOSI{30}{\us}$
The duration of an event is set to be at least as long as the drift length (\SI{2.2}{\ms}).
Correspondingly, for higher level data, the number of items and the data size decreases by orders of magnitude.
For instance, an hour of data may amount to \SI{250}{\giga\byte} of raw-records, \SI{5}{\giga\byte} of peaks, and only \SI{30}{\mega\byte} of events.

The different levels of data processing are organized in software modules called \textit{plugins}, each producing one or more datatypes which can serve as the input data for subsequent (higher level) plugins. 
This structure allows for a modular design and a flexible processing framework. 
When a chunk of data is processed, it is transferred between processing threads to any higher level plugins requiring it as input.
Using this structure, the versioning of the data is handled per datatype by tracking the dependency chain.
This has the benefit for reprocessing that, for example, a new or modified plugin at event level can only require event level input data to (re)compute and does not affect any lower level datatypes.
During processing, auxiliary information on several quantities required for data processing, like PMT gains, are queried from a dedicated collection within the \db{}.
This collection is frequently updated with the latest values to ensure the data are processed with as up to date corrections and detector variables as possible.

\subsection{Online processing}

The processing by strax includes several stages, a full description of all its aspects is beyond the scope of this paper.
Instead, some of the aspects are briefly discussed to illustrate that the full reconstruction of all the data is done live.
The lowest level data, the raw-records, are all written to disk without further processing, allowing  to always go back to the unprocessed data.
After the raw-records level, PMT-traces are baseline-subtracted, inverted and integrated. 
For the TPC, time intervals are obtained wherein photon hits are extracted from the PMT traces to build peak sub-clusters, \textit{peaklets}. 
The peaklets are classified and re-clustered according to their type to obtain peaks; S1-peaks are assumed to consist of only one peaklet, while S2-peaks can consist of many. 
Using this two-step clustering, strax is able to deal with the very short S1 signals while also being able to reconstruct the longer S2 signals as single peaks. 
Three different neural networks are applied on the peak level data for xy-position reconstruction based on the PMT hit pattern, which allows for cross-validation of their results.
Events are built on the basis of a large S2 peak (the ``triggering'' peak). The triggering peak should be >100 PE and there should be fewer than 8 other peaks with at most 50\% of the area of the triggering peak in a \SI{100}{\milli\second} window around the triggering peak. 
An event is the time region from \SI{2.45}{\milli\second} before and \SI{0.25}{\milli\second} after the triggering peak.
This time region is set to be longer than the drift length of \SI{2.2}{\milli\second} (in SR0) and all peaks within the time region are considered part of the event. 
This is effectively the event trigger, which is set as a high-level configuration in the processing chain, in stark contrast to XENON1T~\citep{1t_daq}, where an event window was fixed once at the DAQ and all other data was discarded. 
As a result, the event trigger is easily re-optimized in a high-level analysis.

Processing of \mv{} and \nv{} data is also performed within strax using dedicated veto plugins which are applied similarly to both types of veto data.
These plugins reconstruct veto-events based on the number of PMT hits.
Additionally, a software coincidence trigger for the \nv{} reduces the data stream at a low data level.
This software trigger is not used for the \mv{} because it has a hardware coincidence trigger.

The program \textit{bootstrax} is responsible for processing on the eventbuilders and is optimized per host machine to provide the maximum performance under a wide variety of data rates. 
As soon as a new run is issued by the dedicated program (the dispatcher, see~\autoref{sec:dispatcher}), bootstrax looks for newly written chunks on the Ceph buffer disk.
Bootstrax marks a set of data ready for uploading into long-term storage after completing the processing.

\subsection{Online monitoring\label{sec:olmo}}

\begin{figure}[t]
\includegraphics[width=\textwidth]{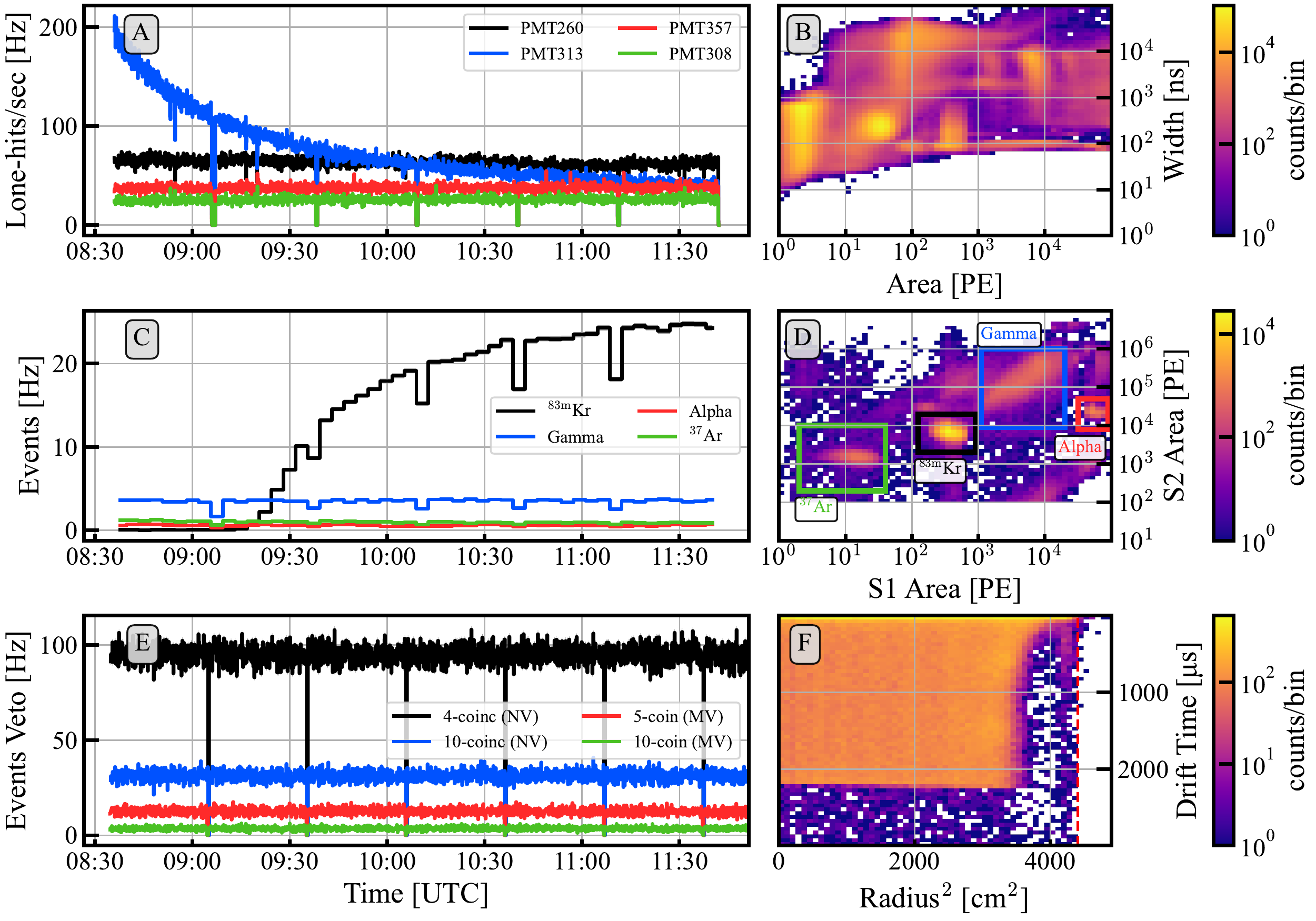}
\caption{Online monitor plot for monitoring the detector status.
    Panel~A shows the per PMT \textit{lone-hit} rate, which are pulses that are seen in one PMT without any pulses in other PMTs within a short time interval. 
    PMT313 (blue) just ``flashed''~\cite{Antochi:2021wik}, and is slowly returning to a rate comparable to the other PMTs. 
    Panel~B shows the area of peaks versus the range of 50 percent decile (known as the width) of the sum-waveform of a peak. 
    This parameter space is useful for identifying peak populations, e.g., the peaks from \iso{Kr}{83m} S1 are visible in the range 80-800 PE at roughly \SI{100}{\ns} width. 
    Panel~C shows the evolution of reconstructed events, which are roughly selected on their S1 and S2 area as shown in figure panel~D. 
    Since this was the start of a calibration period with \iso{Kr}{83m} following an \iso{Ar}{37} calibration, the event rate of \iso{Kr}{83m} is increasing in panel~C while the \iso{Ar}{37} remnants are being removed via online distillation to a negligible level~\citep{10.1093/ptep/ptac074}.
    Panel~E shows the evolution of the number of veto events in the veto-systems over time. 
    Panel~F shows the reconstructed event positions throughout the TPC, where at larger drift times (deeper into the detector), the events are reconstructed inward due to an inhomogeneous drift field inside the detector~\citep{PhysRevD.100.052014}.
    The drops in the rate to \SI{0}{\Hz} (panels~A and E) mark the periods where the DAQ is switching from one run to another.
    As the data in panel~C are re-binned, the run transitions manifest in \order{10\%} drops in the rate. 
}
\label{fig:online_monitor}
\end{figure}
The live processing on the eventbuilders is able to keep up with the data rates observed during SR0, including all the calibration periods. 
This opens up possibilities to use fully reconstructed data to monitor the state of the detector while data collection is ongoing. 
To this end, several datatypes are uploaded while data are being collected. 
This includes acquisition monitor data, all the fully reconstructed events and selections of data from the \mv{}, \nv{}, and a selection of the peaks data from the TPC.

Redax buffers at least two chunks in memory, which first have to be written to the Ceph buffer disk before that data can be processed.
Several chunks are usually combined in memory during processing before writing it to disk to reduce the number of small files in long-term storage. 
However, when chunks of processed data are uploaded to the \db{}, there is no such limitation and a chunk of processed data is therefore immediately uploaded after processing.
This usually results in the data being available in the database $\BigOSI{30}{\second}$ after light has been detected by PMTs.

Status overview plots to monitor the detector conditions and data quality are made with the open-source infrastructure~\cite{strax,straxen} and additional XENONnT software.
As an example, \autoref{fig:online_monitor} shows two unrelated changes in detector conditions close in time: a period of intermittent light emission (``flash'') of PMT313~\cite{Antochi:2021wik} (apparent from panel A) and the start of a calibration period with \iso{Kr}{83m} (most clearly visible in panel C).
This figure can be produced continuously to see changing detector conditions live.
Additionally, each hour a plot is automatically produced and sent to the XENONnT-Slack~\citep{slack} workspace which is used as the common chat room for the entire experiment.
On Slack, one can also easily request periods of time to make this plot for, which is handled automatically by one of the backup eventbuilders. 
Alternatively, the data can be directly retrieved from the \db{} or via strax to do custom analysis, for example to create SuperNova Early Warning System (SNEWS)~\citep{snews} warnings.

\subsection{Data storage infrastructure}

During the commissioning of XENONnT and the first science run (SR0), the DAQ collected \SI{>2}{\peta\byte} of uncompressed data.
To reduce the required amount of long term storage, aggressive compression algorithms like \textit{bz2} (in case of low data rates $\lesssim$\SI[per-mode=symbol]{65}{\mega\byte\per\second}) and \textit{zstd} (in case of higher data rates) are used to compress the low-level data.
The bz2 and zstd algorithms compress the raw data by factors of 5 and 4, respectively.
While this increases CPU usage on the eventbuilders compared to the faster compression algorithms used for high level data, such as \textit{blosc}, CPU usage is usually not the constraining factor for the eventbuilders.

The eventbuilders write to their own hard disks configured in a RAID5 configuration for performance and redundancy resulting in \SI{22}{\tera\byte} of storage per eventbuilder.
These hard disks, shown as local storage in \autoref{fig:overview}, can be accessed by other hosts within the LNGS network.
The data are uploaded from these disks into long-term storage as soon as bootstrax marks them ready for upload in the \db{}.

\section{System Control \& Oversight}
Control and oversight of the DAQ and its associated subsystems are handled via databases, a user-interface website, and a software controller that coordinates the readout processes.
Two additional servers are used in these roles, one in the LNGS surface server room and one underground with the other DAQ servers in the DAQ room.
The surface server hosts the necessary databases for the DAQ, and also acts as a secure gateway through which experts can remotely access the DAQ subnet underground.
The underground server hosts the DAQ website and the software controller.

\subsection{Databases}
Two databases are used for system control, monitoring, and interprocess communication, both implemented in the NoSQL-based MongoDB~\cite{mongo}.
These are referred to as the ``DAQ'' and ``Runs'' databases.
Each database is subdivided into ``collections'', analogous to tables in an SQL-based database, each containing ``documents'' which are analogous to rows.
Unlike SQL-based databases, documents in one collection are not required to have the same schema, which allows for considerable flexibility.

The Runs database is a three-node replica set of servers located in LNGS, University of Chicago, and Rice University, the latter two being the primary XENONnT analysis facilities. 
This ensures that analysts have access to the database in the event of transient network disruption and protects against data loss due to hardware failure.
One collection in this database contains metadata for each discrete run, which includes quantities such as the run start and end times, which of the three detectors were assigned to this run, the full readout configuration of each detector, and a listing of all datatypes for this run and their storage locations.
Another collection contains data produced for online monitoring as described in~\autoref{sec:olmo}.

The DAQ database, in contrast, is expressly for the operation of the readout and not required for analysis.
All data in this database are either only stored temporarily, or change very infrequently and can be restored from periodic backups in the event of data loss.
This database, therefore, is neither replicated nor directly accessible outside of LNGS.
Several collections in this database contain the regular status snapshots used to monitor the various components of the DAQ.
These collections include the status snapshots of the redax readout processes, the health and performance of all the servers and NIM/VME crates in the DAQ system, and the status of the live processing.
These collections are configured with time to live (TTL) indexes to only store data for 3 days, primarily to ensure that queries against these collections remain fast, and also because the information contained can either be reconstructed from the processed data or is additionally written to disk for long-term availability.
Other collections store all available readout configurations, the system operational goal state set by the website, commands being issued to the readout processes, and important logging messages from the various DAQ processes.

\subsection{User interface website}
\begin{figure}[t!]
\includegraphics[width=\textwidth]{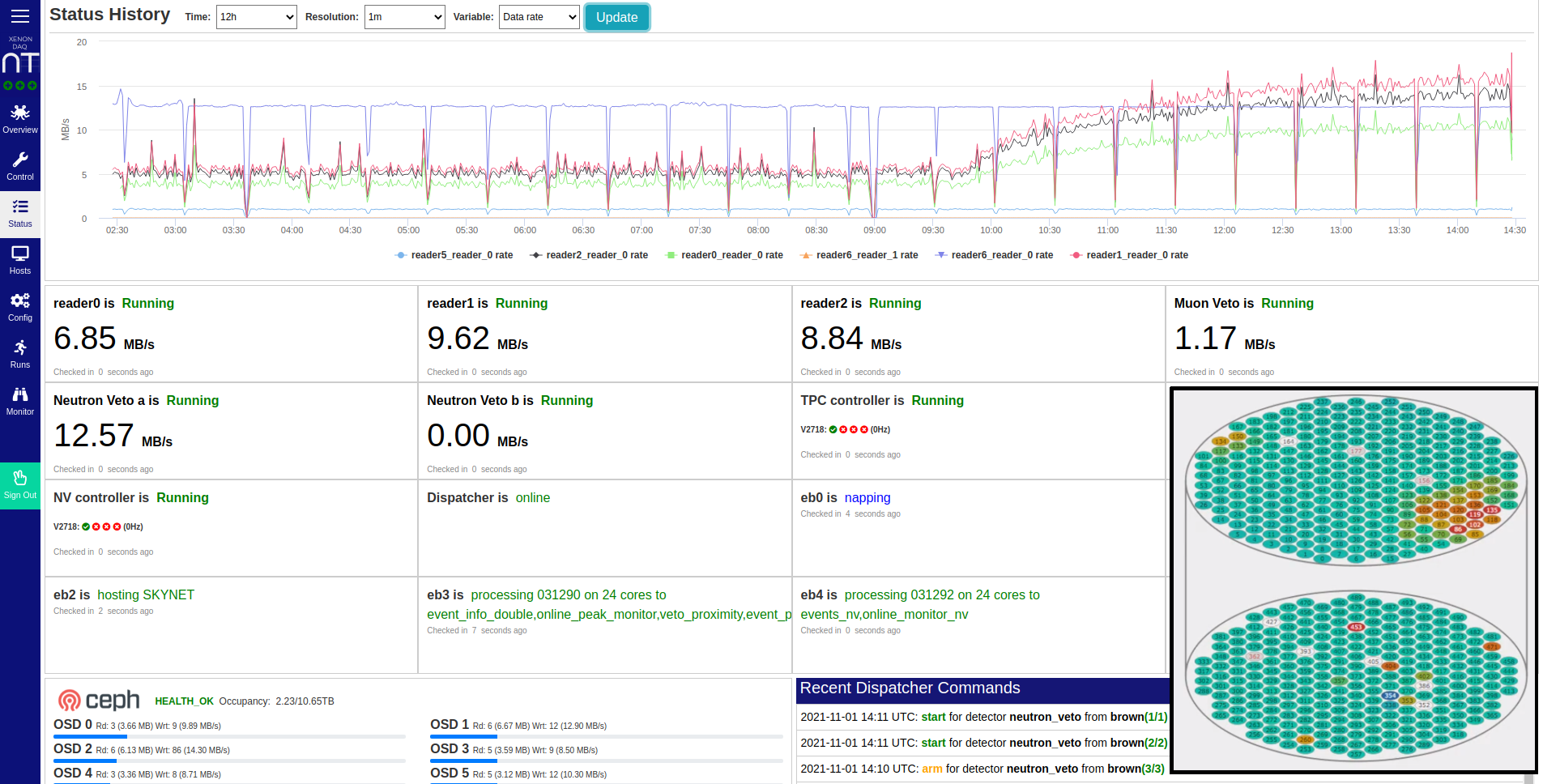}
\caption{
    The Status page on the DAQ interface website. 
    The plot shows the data rates for the past 12 hours, and the status cards show the instantaneous statuses of all readout and live processing elements in the DAQ system. 
    An increase in the rate due to the start of regular detector calibration with \iso{Kr}{83m} is clearly visible. 
    A navigation bar on the left provides convenient links to other pages in the website. 
    The inset in the bottom right shows the per-channel rate as displayed on the monitor page.}
\label{fig:website_status}
\end{figure}

To facilitate easy use of the DAQ for the day-to-day operation of the experiment, a front-end website was developed using NodeJS~\cite{nodejs}.
A variety of pages allows users to view the current readout performance, set the system's operation goal state, and monitor the data rates from each readout channel to identify potential problems in the detector, such as localized regions of sustained electron emission, also known as ``hotspots''.
The \textit{status} page displays instantaneous data rates for each readout process in the entire DAQ system, information about the current activity of each eventbuilder, and the current status of the Ceph high-speed buffer disk.
Additionally, a plot displays the recent data rate for each readout process, which allows a user to identify transient behavior in the system that may not be clear from the instantaneous rates alone.
This page is shown in \autoref{fig:website_status}.
The \textit{control} page allows users to specify the operational goal state of all three subsystems, such as selecting an operational mode and a desired run duration.
The \textit{monitor} page displays the instantaneous data rate for each channel in the TPC in a convenient layout mirroring that of the physical locations of PMTs in the TPC (shown as the inset in \autoref{fig:website_status}).
In this view, a hotspot will appear as a localized region, typically three adjacent PMTs, with a data rate significantly higher than other nearby PMTs.
Additionally, a plotting function is provided to allow for a direct comparison of recent rates between different channels.
Another page provides a user interface to the Runs database, where metadata about each run can be viewed.
Other pages allow experts to modify and create preset operational modes and configurations, monitor the status of all the servers in the DAQ network, and interface with the dispatcher (described below).

Finally, an application programming interface (API) is provided to enable programmatic control of the DAQ and access to part of the DAQ database.
This is used, for instance, by Slow Control to perform the periodic automatic calibration of the PMTs via a pulsed LED.
Slow Control continuously queries the API, and notifies experts if any aspects of the performance of the system deviate from what is expected, if disks are full, or a hotspot is suspected based on the per channel data rate.

\subsection{Readout coordination software\label{sec:dispatcher}}
To oversee and coordinate the readout processes, a program called the \textit{dispatcher} was developed.
The primary responsibility of the dispatcher is to convert the desired operational goal state as specified on the website into direct commands issued to the various readout processes.
To do this, the dispatcher retrieves the most recent status snapshot of each readout process.
These are aggregated together to determine an overall status for each subsystem, such as if the subsystem is idle, running, in a transitional state, or if processes are not responding.
This aggregated status is then compared to the desired operational goal state of each subsystem, and commands are issued to the readout processes to make the former match the latter.
For example, if a user wants the readout to begin with a certain operational mode, the dispatcher will ensure that the necessary processes are capable of starting, issue commands to begin the digitizer-programming sequence, wait until all necessary processes report the successful completion of this sequence, and then issue the start command.
When an active run reaches the desired length as specified by the website, the readout is stopped, and the cycle is repeated.
In rare cases where a readout process or digitizer stops responding properly, the dispatcher will automatically kill and restart delinquent readout processes and power-cycle VME crates as necessary to restore normal behavior.
Additionally, if such action is necessary during linked-mode operation and restarting a process or VME crate fails to rectify the situation, the dispatcher will unlink the detectors so the readout of the detectors that are responding normally can continue.
Experts receive notifications whenever automatic actions such as these are taken.
At the start of every run, the dispatcher creates an entry in the Runs database containing a copy of the readout configuration and other metadata necessary for the live processing.
At the end of a run, the corresponding entry is updated with additional quantities such as the end time of the run and the average data rates for all contributing detectors.

\section{Performance}
In the first two years of operation, the TPC subsystem collected more than \SI{1280}{\tera\byte} of data, while the \mv{} and \nv{} collected \SI{28}{\tera\byte} and \SI{680}{\tera\byte}, respectively.
The performance of the overall DAQ system can be measured in several ways.
Most obvious are the maximum data rate the system can maintain and the livetime with which the system operates, but other criteria such as inter-detector synchronization and noise levels are also important.
An additional key performance metric is the speed of the live processing, as it is crucial that the data are made available for transfer off-site at least as fast as it is recorded.

\subsection{Noise levels}
Externally triggered, short runs were taken weekly throughout SR0 to assess the noise levels in all TPC channels, using fixed windows of approximately \SI{1}{\ms} duration.
The mean RMS noise level was found to be stable for each PMT array with values of \SI{0.23}{\mV} and \SI{0.34}{\mV} for the high-gain channels in the top and bottom arrays, respectively, and \SI{0.16}{\mV} for the low-gain channels in the top array.
The installed filter boxes are effective in suppressing electronic noise \SI{>250}{\kHz}, which is related to HV power supplies.
However, as expected, the filter boxes have a negligible effect on the low-frequency noise peak at \SI{24.41}{\kilo\hertz}, which is correlated with intrinsic noise produced by CAEN digitizers.
Channels from the bottom PMT array on average exhibited an RMS noise level which was $\sim$1 ADCc higher when compared to top PMT array channels.
This effect could be attributed to either a different resistor type and assembly procedure that was employed for filter boxes used for the bottom PMT array, or different and noisier HV power supplies.
These low levels of noise support low digitization thresholds for the PMTs.
Over $98\%$ of TPC channels have thresholds set at \SI{2}{\mV} and only 1 PMT has a threshold higher than \SI{3.4}{\mV}, giving an average acceptance to single PE $>90\%$.
For the \nv, 109 PMTs ($91\%$) have thresholds set at \SI{1.8}{\mV} ($\sim0.3\,\mathrm{PE}$) and the other 11 at \SI{2.4}{\mV} ($\sim0.4\,\mathrm{PE}$).
The average noise for the \nv{} PMTs is \SI{0.3}{\milli\volt} (\SI{\sim0.05}{\pe}).
Lastly, for the \mv{} the average RMS noise is \SI{0.18}{\mV}, the thresholds are set for all channels at \SI{3}{\mV} ($\sim1\,\mathrm{PE}$).
All above voltages were estimated for an input impedance of \SI{50}{\ohm}.

\subsection{Livetime \label{sec:deadtime}}

\begin{figure}[t!]
\includegraphics[width=\textwidth]{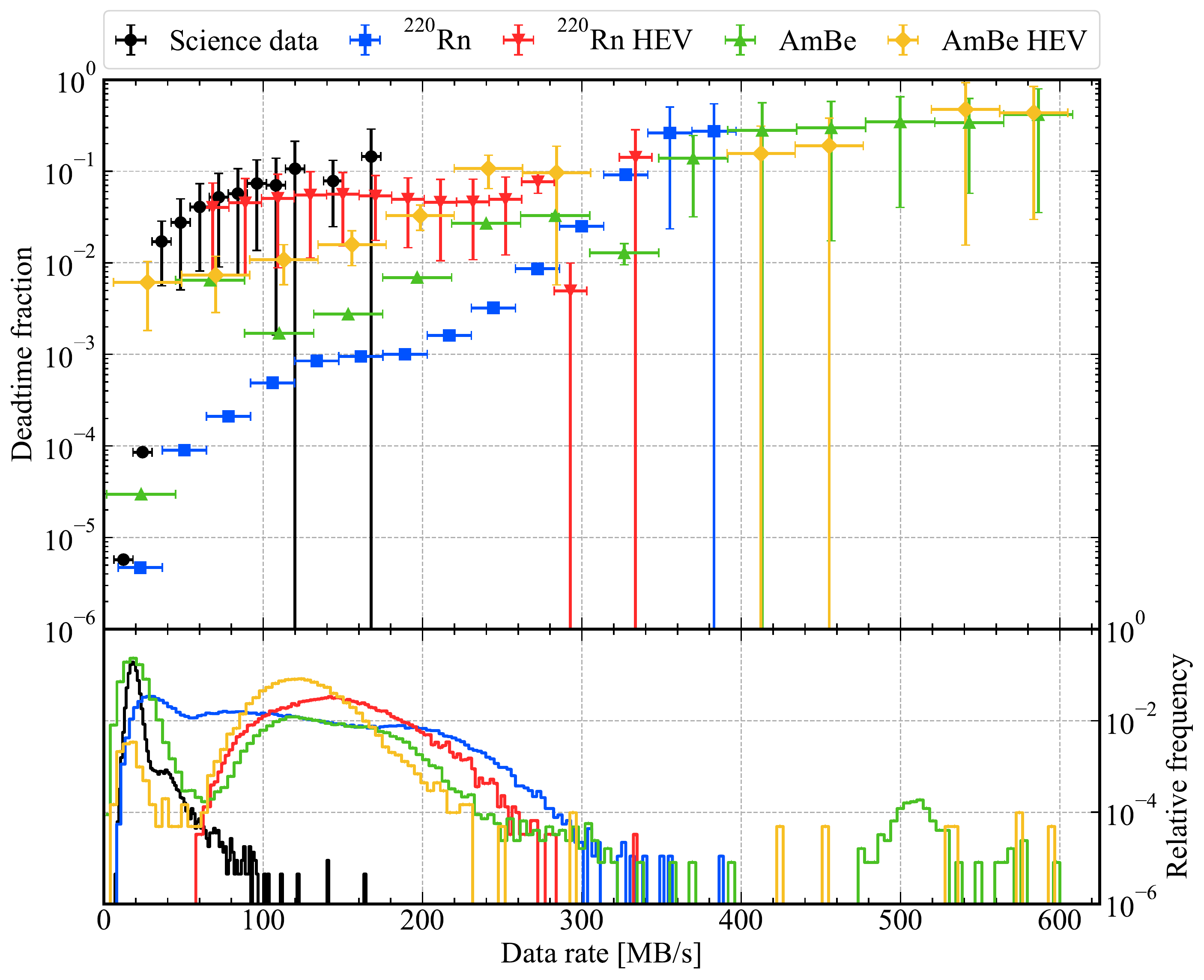}
\caption{
    Mean total deadtime per chunk (top panel), and relative frequency of chunks (bottom panel) as a function of data rate per chunk for several run modes.
    Each chunk is a time interval of \SIrange{5}{20}{\second}.
    AmBe calibrations are performed by keeping the source at several positions with respect to the TPC, leading to distinct populations in the bottom panel.
    The High Energy Veto (HEV) run modes (AmBe HEV and \iso{Rn}{220} HEV) have higher deadtime fractions as a result of the inserted deadtime by the HEV, see \autoref{sec:busyhev}.
    For typical (98\%) science data ($\lesssim$\SI[per-mode=symbol]{25}{\mega\byte\per\second}) the deadtime fraction is \SI{2e-05}{}.
    In science data, higher rate data points are caused by short periods in time following a muon traversing the TPC, leading to high data rates and deadtime fractions of $\mathcal{O}(1\%)$.
    \iso{Rn}{220} has a lower deadtime fraction for $\gtrsim$\SI[per-mode=symbol]{25}{\mega\byte\per\second} than science data, since these higher rates are caused by a higher S2 rate, rather than muons.
    Above \SI[per-mode=symbol]{\sim250}{\mega\byte\per\second} the data quality deteriorates due to the onset of pileup.
}
\label{fig:deadtime}
\end{figure}

Throughout SR0 and the commissioning of XENONnT all three DAQ subsystems operated stably, collecting in total more than 200 days of commissioning and science data, and close to 100 days of various calibration data.
The deadtime fraction induced by the operation of the busy veto is  \SI{2e-5}{} for the majority of SR0 science data (which is typically $\lesssim$\SI[per-mode=symbol]{25}{\mega\byte\per\second}), as illustrated in \autoref{fig:deadtime}. 
The average deadtime fraction for all SR0 science data is \SI{3e-4}{}.
Furthermore, during high-rate \iso{Rn}{220} and AmBe calibration periods the deadtime resulting from the combined operation of the busy and HEV on average amounts to $\sim10\%$. 
The above deadtime values describe only the intrinsic deadtime produced by the operation of the busy and HEV modules, and not the data reduction caused by the analysis cut described in \autoref{sec:acqmon}. 
Lastly, it should be noted that the busy veto-induced deadtime of the NV was found to be negligible.

The highest sustained data rate in SR0 was \SI[per-mode=symbol]{\sim500}{\mega\byte\per\second} during an AmBe calibration (population near \SI[per-mode=symbol]{500}{\mega\byte\per\second} in the bottom panel of \autoref{fig:deadtime}).
The DAQ was designed to withstand higher data rates, but performing high rate calibrations in SR0 was inhibited by the long event duration which leads to pileup of events where they start overlapping.
A few chunks were digitized with data rates of up to \SI[per-mode=symbol]{600}{\mega\byte\per\second}.

\subsection{Time synchronization} 
Throughout SR0 data-taking the three DAQ subsystems operated in linked mode with full temporal synchronization.
However, as was described above, to facilitate the operation of the HEV the TPC data are delayed within the V1724 digitizers by \SI{10}{\us}, as compared to the TPC acquisition monitor and the other sub-detectors.
The time synchronization across all sub-detectors was verified using the \SI{0.1}{\hertz} signal generated by the GPS. 
It was supplied to a dedicated TPC digitizer channel, as well as to its acquisition monitor.
Additionally, this signal was acquired by the acquisition monitor of the \nv{} and a digitizer in the \mv{}.
A comparison between the timestamps of these signals was used to measure the average temporal difference between the sub-detectors.
The average time difference between the TPC and \nv{} was measured to be \SI{10157}{\ns}, while the time difference between the TPC and \mv{} was found to be \SI{5283}{\ns}. 
The variation in the measured delay time is related to the trigger formation time used by the \mv{}.
After SR0 these constant offsets are subtracted out during readout, resulting in time synchronization with a precision of \SI{\sim 10}{\ns}, which is comparable to the sampling time of the digitizers.

\begin{figure}[!t]
\includegraphics[width=1\textwidth]{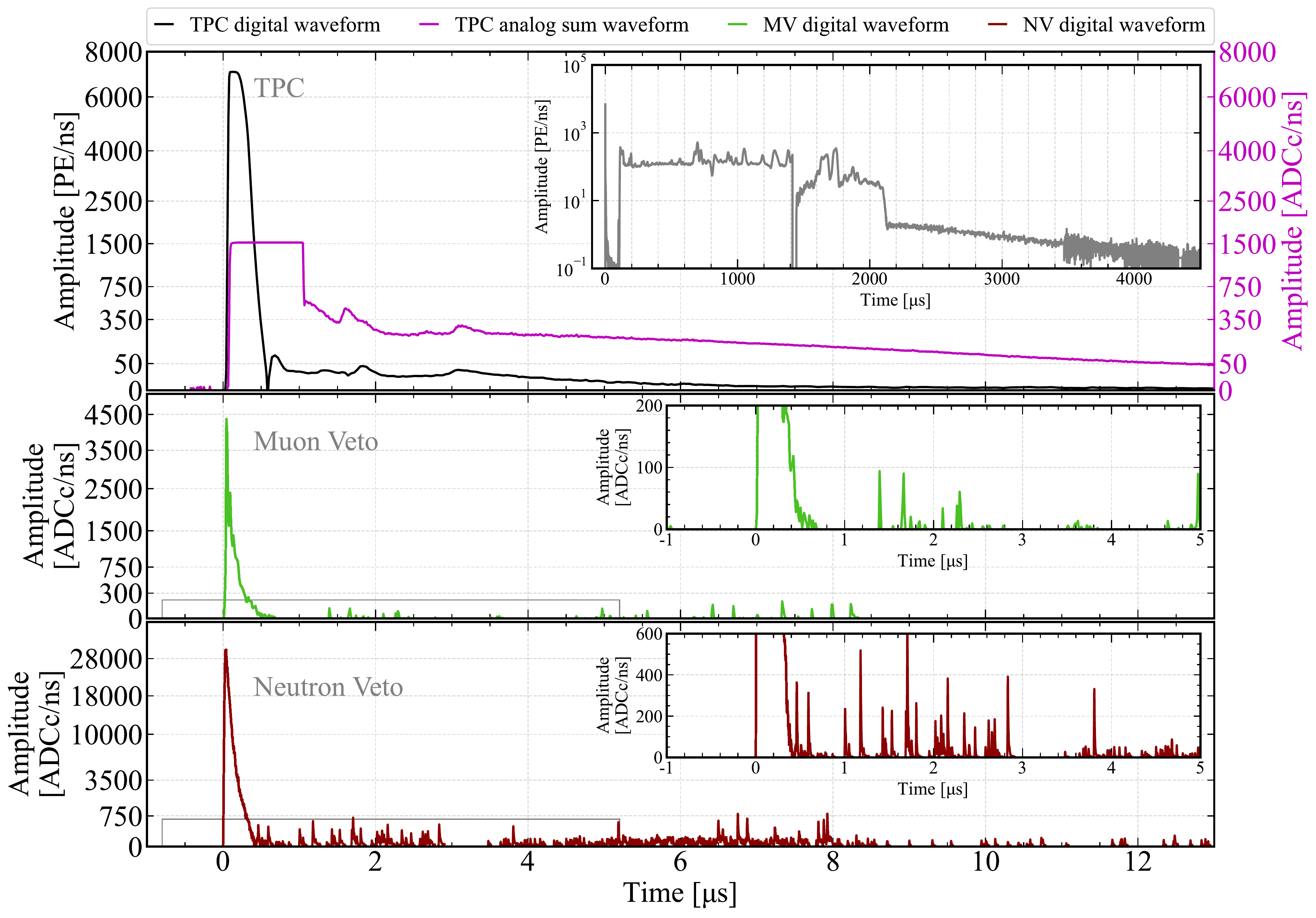}
\caption{
    An illustration of a muon event recorded by all three DAQ subsystems. 
    The top panel shows a zoomed-in view of the beginning of the muon waveform that was recorded by the TPC digitizers (black). 
    An analog sum waveform of the bottom PMT array that was recorded by the TPC's acquisition monitor digitizer is also shown in the same panel (magenta). 
    The inset in the top panel shows the entire muon waveform duration as seen by the TPC digitizers.
    The drop at \SI{\sim 1400}{\us} is caused by a baseline fluctuation.
    The middle panel shows the same muon event recorded by the \mv{}, while the bottom panel shows the data recorded by the \nv{}. 
    Insets in both middle and bottom panels show a zoomed-in view of the respective waveforms. 
    All three waveforms were aligned based on the TPC signal (black).}
\label{fig:muon_event}
\end{figure}

To illustrate the inter-detector synchronization a muon event passing through all XENONnT sub-detectors is presented in \autoref{fig:muon_event}. 
The signals obtained in each detector are aligned by accounting in software for the time differences described above.
In addition to the sub-detector signals, also the analog sum waveform that is acquired by the acquisition monitor of the TPC is shown in the top panel. 
As seen in the inset of the top panel of \autoref{fig:muon_event}, the prompt S1 from the relativistic muon's interactions is followed by a sustained S2, which lasts for the full drift time of the detector. This extended S2 indicates a vertically traversing muon interacting along most of the drift column to produce ionization electrons. The subsequent long tail is formed from photoionization electrons that follow the S2 for multiple milliseconds.
The muon track in the \mv{} and \nv{} sub-detectors corresponds to the Cherenkov photons detected by the PMTs of these detectors, followed by PMT afterpulses lasting up to \SI{\sim 10}{\us}. 
The general shape of the waveform in all three panels is the same, indicating clearly that the same event is shown.
Lastly, the GPS synchronization signal was also used to evaluate the clock drift of the DT4700 clock module, yielding approximately \SI{2}{\us} over a period of \SI{10}{\s}, or $0.2\,\mathrm{ppm}$.

\subsection{Live processing performance}
The DAQ and its software were designed to run under high data rates during detector calibrations using a combination of internal and external radioactive sources to quantify the detector performance. 
However, due to a limited voltage on the TPC cathode during SR0 compared to the design value, high rate calibrations could not be performed as events quickly piled up because of the long drift time of \SI{2.2}{\milli\second}.
As a consequence, the system did not have to work under persistent high data rates and live processing was always able to process the data faster than it could be collected. 
During SR0, the data rate never exceeded $\sim$\SI[per-mode=symbol]{50}{\mega\byte\per\second} for extended periods of time.
To quantify the performance of the eventbuilders in high data rate conditions, pre-SR0 commissioning data that were taken during a high rate \iso{Kr}{83m} calibration are used.

\begin{figure}[htb]
\includegraphics[width=\textwidth]{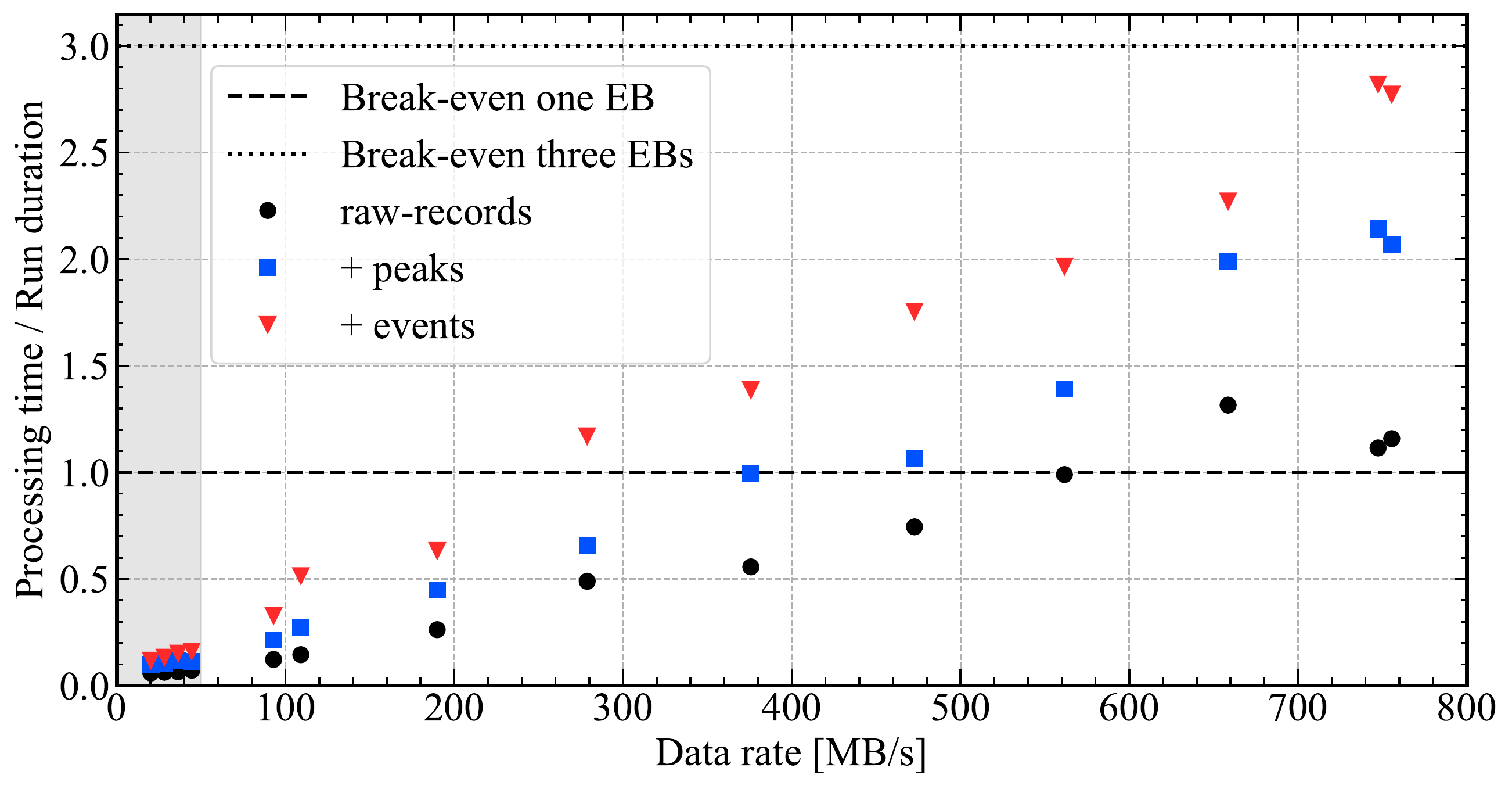}
\caption{
    Live processing time as a function of the raw data rate for several target datatypes. 
    The raw-records datatype is the lowest level datatype, followed by peaks and finally events (for simplicity this is called events, even though the benchmarks were obtained for the event-basics datatype \citep{straxendocumentation}). 
    When a higher level datatype is computed, all the lower level datatypes are also produced, so processing event also includes raw-records and peaks. 
    All SR0 science data in this figure are \SI[per-mode=symbol]{50}{\mega\byte\per\second} (gray band).
    Pre-SR0 commissioning data were used for the high rate data points where the DAQ was operated with a fractional livetime mode (discussed in \autoref{sec:busyhev}) during a high rate \iso{Kr}{83m}-calibration.
    For data rates $\lesssim$\SI[per-mode=symbol]{250}{\mega\byte\per\second} the live processing keeps up with one eventbuilder (EB) server as the processing time is lower than the acquisition time for any datatype. 
    For any data rate in this plot the points are below the break-even line of the three eventbuilders, meaning that live processing at the DAQ could keep up with the readout.
}
\label{fig:strax_rate}
\end{figure}

\autoref{fig:strax_rate} shows bootstrax cumulative processing time for different datatypes. 
Here, raw-records is the lowest level datatype, followed by peaks and finally events.
The results for higher level datatypes include the time to compute the lower level datatypes as well.
There are some additional datatypes of intermediate data in between, which one can find in the straxen documentation~\citep{straxendocumentation}, several of which were briefly discussed in \autoref{sec:live_processing}. 
The total processing time comprises the time of starting bootstrax for a given run,
decompressing the redax data, processing the data until the specified datatypes, compressing and writing all of the processed data to disk. 

\autoref{fig:strax_rate} shows that for data rates below \SI[per-mode=symbol]{250}{\mega\byte\per\second}, the processing time is shorter than the collection time and a single eventbuilder can manage the entire data stream regardless of the datatype considered. 
For higher rates, the processing up to the events or peaks datatype is not fast enough to keep the processing live as each chunk would be processed slightly later than it is acquired. 
At these data rates, the finite RAM of the servers and increased disk read/write operations prevent processing at the same rate as at lower data rates, since processing each new chunk on a separate core starts requiring more memory than available on the host.
The break-even line for one eventbuilder in \autoref{fig:strax_rate} lies around \SI[per-mode=symbol]{\sim 250}{\mega\byte\per\second}  for events, \SI[per-mode=symbol]{\sim 400}{\mega\byte\per\second} for peaks, and \SI[per-mode=symbol]{\sim 550}{\mega\byte\per\second} for raw-records. 
Additionally, the work is divided among three eventbuilders (with two additional as backup) and the combined eventbuilders can keep up with much higher data rates. 

The total rate of the \nv{} and \mv{} subsystems was found to be relatively constant and amounts to \SIrange[per-mode=symbol,range-phrase=--]{10}{20}{\mega\byte\per\second} and \SI[per-mode=symbol]{\sim1}{\mega\byte\per\second}, respectively. 
It takes about \SI{80}{\second} to process a \SI{1800}{\second} run for the \nv~data, and $\lesssim$\SI{40}{\second} for a \SI{1800}{\second} run of \mv~data.

\section{Conclusion}
The XENON collaboration has designed and commissioned the triggerless XENONnT DAQ. 
By forgoing a trigger and relying instead on fast software to handle the continuous data stream, all data exceeding the digitization threshold is written to disk. 
The TPC, \mv{} and \nv{} subsystems that constitute the DAQ can be operated independently, or as one linked system sharing the same \SI{50}{\MHz} clock signal.
The increased number of PMTs and the double digitization of the top PMT array leads to roughly three times the number of channels with respect to XENON1T for the TPC.
While the triggered \mv{}-subsystem remains virtually unchanged, the new \nv{}-subsystem was successfully built to enable tagging of neutron events, one of the main backgrounds for the XENONnT WIMP search.
A \SI{500}{\mega\hertz} sampling rate enables the required characterization of neutron signals.

The DAQ is able to operate at the highest data rates observed during the first science run of XENONnT (SR0) of \SI[per-mode=symbol]{\sim500}{\mega\byte\per\second} with the potential to go higher.
The deadtime fraction is as low as \SI{3e-4}{} for science data and $\lesssim10\%$ for calibration data at high rates of \SI[per-mode=symbol]{<350}{\mega\byte\per\second}.

Using online processing, high level data are directly made available to monitor the detector.
This enables analysts to have immediate feedback on changing detector conditions with fully processed data.
The online processing is able to handle all data rates observed during SR0, where each of the three dedicated servers is able to process the data with a rate of up to \SI[per-mode=symbol]{\sim250}{\mega\byte\per\second}.
The maximum observed data rates during SR0 were limited by the low drift-field conditions. 
However, the DAQ was designed and is capable of dealing with data throughput rates greater than \SI[per-mode=symbol]{750}{\mega\byte\per\second}.

During commissioning and SR0, the XENONnT DAQ has collected more than \SI{2}{\peta\byte} of both science and calibration data, and it will continue to operate in subsequent science runs. 
The successful operation of the XENONnT DAQ and the implementation of the triggerless readout paradigm provides a solid basis for the development of DAQ systems for the next generation of liquid xenon dark matter experiments~\citep{darwin, darwin_design}.

\acknowledgments
We gratefully acknowledge support from the National Science Foundation, Swiss National Science Foundation, German Ministry for Education and Research, Max Planck Gesellschaft, Deutsche Forschungsgemeinschaft, Helmholtz Association, Dutch Research Council (NWO), Weizmann Institute of Science, Israeli Science Foundation, Binational Science Foundation, Fundacao para a Ciencia e a Tecnologia, R\'egion des Pays de la Loire, Knut and Alice Wallenberg Foundation, Kavli Foundation, JSPS Kakenhi and JST FOREST Program in Japan, Tsinghua University Initiative Scientific Research Program and Istituto Nazionale di Fisica Nucleare. This project has received funding/support from the European Union’s Horizon 2020 research and innovation programme under the Marie Sk\l{}odowska-Curie grant agreement No 860881-HIDDeN. Data processing is performed using infrastructures from the Open Science Grid, the European Grid Initiative and the Dutch national e-infrastructure with the support of SURF Cooperative. We are grateful to Laboratori Nazionali del Gran Sasso for hosting and supporting the XENON project. 

\bibliographystyle{JHEP}
\bibliography{biblio.bib}

\providecommand{\href}[2]{#2}\begingroup\raggedright\begin{thebibliography}{10}

\bibitem{1t_instrument}
{\scshape XENON} collaboration, \emph{The {XENON1T} dark matter experiment},
  \href{https://doi.org/10.1140/epjc/s10052-017-5326-3}{\emph{Eur. Phys. J. C}
  {\bfseries 77} (2017) 881}.

\bibitem{lz}
{\scshape LZ} collaboration, \emph{{The LUX-ZEPLIN (LZ) Experiment}},
  \href{https://doi.org/10.1016/j.nima.2019.163047}{\emph{Nucl. Instrum. Meth.
  A} {\bfseries 953} (2020) 163047}
  [\href{https://arxiv.org/abs/1910.09124}{{\ttfamily 1910.09124}}].

\bibitem{pandax}
{\scshape PandaX-4T} collaboration, \emph{{Dark Matter Search Results from the
  PandaX-4T Commissioning Run}},
  \href{https://doi.org/10.1103/PhysRevLett.127.261802}{\emph{Phys. Rev. Lett.}
  {\bfseries 127} (2021) 261802}
  [\href{https://arxiv.org/abs/2107.13438}{{\ttfamily 2107.13438}}].

\bibitem{nexo}
{\scshape nEXO} collaboration, \emph{{nEXO: neutrinoless double beta decay
  search beyond 10$^{28}$ year half-life sensitivity}},
  \href{https://doi.org/10.1088/1361-6471/ac3631}{\emph{J. Phys. G} {\bfseries
  49} (2022) 015104} [\href{https://arxiv.org/abs/2106.16243}{{\ttfamily
  2106.16243}}].

\bibitem{darkside}
{\scshape DarkSide} collaboration, \emph{{DarkSide-50 532-day Dark Matter
  Search with Low-Radioactivity Argon}},
  \href{https://doi.org/10.1103/PhysRevD.98.102006}{\emph{Phys. Rev. D}
  {\bfseries 98} (2018) 102006}
  [\href{https://arxiv.org/abs/1802.07198}{{\ttfamily 1802.07198}}].

\bibitem{1tonneyear}
{\scshape XENON} collaboration, \emph{{Dark Matter Search Results from a One
  Ton-Year Exposure of XENON1T}},
  \href{https://doi.org/10.1103/PhysRevLett.121.111302}{\emph{Phys. Rev. Lett.}
  {\bfseries 121} (2018) 111302}
  [\href{https://arxiv.org/abs/1805.12562}{{\ttfamily 1805.12562}}].

\bibitem{nt_mc}
{\scshape XENON} collaboration, \emph{{Projected WIMP sensitivity of the
  XENONnT dark matter experiment}},
  \href{https://doi.org/10.1088/1475-7516/2020/11/031}{\emph{JCAP} {\bfseries
  11} (2020) 031} [\href{https://arxiv.org/abs/2007.08796}{{\ttfamily
  2007.08796}}].

\bibitem{low_er_nt}
{\scshape XENON} collaboration, \emph{Search for new physics in electronic
  recoil data from xenonnt},
  \href{https://doi.org/10.1103/PhysRevLett.129.161805}{\emph{Phys. Rev. Lett.}
  {\bfseries 129} (2022) 161805}.

\bibitem{caen_general}
``Caen.'' \url{https://www.caen.it/}.

\bibitem{1t_daq}
{\scshape XENON} collaboration, \emph{The {XENON1T} data acquisition system},
  \href{https://doi.org/10.1088/1748-0221/14/07/p07016}{\emph{JINST} {\bfseries
  14} (2019) P07016}.

\bibitem{mv_paper}
{\scshape XENON} collaboration, \emph{{Conceptual design and simulation of a
  water Cherenkov muon veto for the XENON1T experiment}},
  \href{https://doi.org/10.1088/1748-0221/9/11/P11006}{\emph{JINST} {\bfseries
  9} (2014) P11006} [\href{https://arxiv.org/abs/1406.2374}{{\ttfamily
  1406.2374}}].

\bibitem{julien_amp_thesis}
J.~Wulf, \emph{Direct Dark Matter Search with XENON1T and Developments for
  Multi-Ton Liquid Xenon Detectors}, Ph.D. thesis, Universität Zürich, 2018.

\bibitem{mongo}
``Mongodb.'' \url{https://mongodb.com}.

\bibitem{pax}
J.~Aalbers, C.~Tunnell, B.~Pelssers, A.~Buß, F.~Gao, Q.~Lin et~al., ``{PAX}
  data processor.'' \url{https://doi.org/10.5281/zenodo.4290785}, Mar., 2020.
\newblock 10.5281/zenodo.4290785.

\bibitem{strax}
J.~Aalbers, J.R.~Angevaare, D.~Wenz, C.~Tunnell, Y.~Mosbacher, P.~Gaemers
  et~al., ``Axfoundation/strax: v1.1.5.''
  \url{https://doi.org/10.5281/zenodo.5833114}, Jan., 2022.
\newblock 10.5281/zenodo.5833114.

\bibitem{straxen}
J.R.~Angevaare, J.~Aalbers, D.~Wenz, E.~Shockley, P.~Gaemers, A.~Higuera
  et~al., ``Xenonnt/straxen: v1.2.4.''
  \url{https://doi.org/10.5281/zenodo.5834311}, Jan., 2022.
\newblock 10.5281/zenodo.5834311.

\bibitem{numpy}
C.R.~Harris, K.J.~Millman, S.J.~van~der Walt, R.~Gommers, P.~Virtanen,
  D.~Cournapeau et~al., \emph{Array programming with {NumPy}},
  \href{https://doi.org/10.1038/s41586-020-2649-2}{\emph{Nature} {\bfseries
  585} (2020) 357}.

\bibitem{scipy}
P.~Virtanen, R.~Gommers, T.E.~Oliphant, M.~Haberland, T.~Reddy, D.~Cournapeau
  et~al., \emph{{{SciPy} 1.0: Fundamental Algorithms for Scientific Computing
  in Python}}, \href{https://doi.org/10.1038/s41592-019-0686-2}{\emph{Nature
  Methods} {\bfseries 17} (2020) 261}.

\bibitem{numba}
S.K.~Lam, A.~Pitrou and S.~Seibert, \emph{Numba: A llvm-based python jit
  compiler},  in \emph{Proceedings of the Second Workshop on the LLVM Compiler
  Infrastructure in HPC}, pp.~1--6, 2015.

\bibitem{gps_slave_master}
M.~De~Deo et~al., \emph{Accurate {GPS}-based timestamp facility for {G}ran
  {S}asso {N}ational {L}aboratory},
  \href{https://doi.org/10.1088/1748-0221/14/04/p04001}{\emph{JINST} {\bfseries
  14} (2019) P04001}.

\bibitem{ddc10}
SkuTek Instrumentation, \emph{Digital Pulse Processor for Nuclear Physics}.
\newblock \url{https://www.skutek.com/ddc10.htm}.

\bibitem{ceph}
``Ceph.'' \url{https://ceph.io}.

\bibitem{redax}
D.~Masson, D.~Coderre, J.R.~Angevaare, A.~Elykov, S.D.~Pede, P.~Gaemers et~al.,
  ``Axfoundation/redax: Version 2.3.0.''
  \url{https://doi.org/10.5281/zenodo.5882717}, Jan., 2022.
\newblock 10.5281/zenodo.5882717.

\bibitem{Antochi:2021wik}
V.C.~Antochi et~al., \emph{{Improved quality tests of R11410-21 photomultiplier
  tubes for the XENONnT experiment}},
  \href{https://doi.org/10.1088/1748-0221/16/08/P08033}{\emph{JINST} {\bfseries
  16} (2021) P08033} [\href{https://arxiv.org/abs/2104.15051}{{\ttfamily
  2104.15051}}].

\bibitem{10.1093/ptep/ptac074}
{\scshape XENON} collaboration, \emph{{Application and modeling of an online
  distillation method to reduce krypton and argon in XENON1T}},
  \href{https://doi.org/10.1093/ptep/ptac074}{\emph{PTEP} {\bfseries 2022}
  (2022) 053H01} [\href{https://arxiv.org/abs/2112.12231}{{\ttfamily
  2112.12231}}].

\bibitem{PhysRevD.100.052014}
{\scshape XENON} collaboration, \emph{Xenon1t dark matter data analysis: Signal
  reconstruction, calibration, and event selection},
  \href{https://doi.org/10.1103/PhysRevD.100.052014}{\emph{Phys. Rev. D}
  {\bfseries 100} (2019) 052014}
  [\href{https://arxiv.org/abs/1906.04717}{{\ttfamily 1906.04717}}].

\bibitem{slack}
``Slack.'' \url{https://slack.com/}.

\bibitem{snews}
S.~Al~Kharusi, S.Y.~BenZvi, J.S.~Bobowski, W.~Bonivento, V.~Brdar, T.~Brunner
  et~al., \emph{{SNEWS} 2.0: a next-generation supernova early warning system
  for multi-messenger astronomy},
  \href{https://doi.org/10.1088/1367-2630/abde33}{\emph{New Journal of Physics}
  {\bfseries 23} (2021) 031201}.

\bibitem{nodejs}
``Nodejs.'' \url{https://nodejs.dev}.

\bibitem{straxendocumentation}
``Straxen documentation.'' \url{https://straxen.readthedocs.io/}.

\bibitem{darwin}
J.~Aalbers et~al., \emph{{A Next-Generation Liquid Xenon Observatory for Dark
  Matter and Neutrino Physics}},
  \href{https://doi.org/https://doi.org/10.48550/arXiv.2203.02309}{\emph{arXiv}
  (2022) } [\href{https://arxiv.org/abs/2203.02309}{{\ttfamily 2203.02309}}].

\bibitem{darwin_design}
{\scshape DARWIN} collaboration, \emph{Darwin: towards the ultimate dark matter
  detector}, \href{https://doi.org/10.1088/1475-7516/2016/11/017}{\emph{Journal
  of Cosmology and Astroparticle Physics} {\bfseries 2016} (2016) 017}.

\end{thebibliography}\endgroup

\end{document}